\documentclass[sigconf]{acmart}

\usepackage{booktabs} %
\usepackage{tikz,subfig} 
\usepackage{color,enumitem}
\usepackage{listings}             %
\usepackage{times}
\usepackage{balance}

\newcommand{\hpara}[1]{\textcolor{blue}{[#1]}}
\renewcommand{\hpara}[1]{}

\newcommand{\abbr}{\texorpdfstring{\MakeLowercase{s}}{s}PIN}

\setcopyright{none}
\acmConference[SC'17]{ACM/IEEE The International Conference for High
Performance Computing, Networking, Storage and Analysis}{Nov. 2017}{Denver, CO, USA} 
\copyrightyear{2017}
\acmYear{2017}
\setcopyright{usgovmixed}
\acmConference{SC17}{November 12--17, 2017}{Denver, CO, USA}\acmPrice{15.00}\acmDOI{10.1145/3126908.3126970} \acmISBN{978-1-4503-5114-0/17/11}

\begin{document}
\title{sPIN: High-performance \underline{s}treaming \underline{P}rocessing \underline{i}n the \underline{N}etwork}
%sPIN: High-performance streaming Processing in the Network
%
%
%
%

\author{Torsten Hoefler$^{\dagger\ast}$, Salvatore Di
Girolamo$^\dagger$, Konstantin Taranov$^\dagger$, Ryan E.
Grant$^\ast$, Ron Brightwell}
\authornote{Sandia National Laboratories is a multimission laboratory managed 
and operated by National Technology and Engineering Solutions of Sandia, LLC., 
a wholly owned subsidiary of Honeywell International, Inc., for the U.S. 
Department of Energy's National Nuclear Security Administration under contract 
DE-NA-0003525.}
\affiliation{%
  \begin{tabular}{p{.3\textwidth}p{.3\textwidth}}
    \centering
  \institution{$^\dagger$ETH Z\"urich}
  \streetaddress{Universit\"atsstrasse 6}
  \city{\ \ \ \ 8092 Z\"urich, Switzerland}\newline 
{\{htor,digirols,ktaranov\}@inf.ethz.ch}
  &
    \centering
  \institution{$^\ast$Sandia National Laboratories}
  \streetaddress{P.O. Box 5800}
  \city{\ \ \ \ Albuquerque, NM, USA}\newline
{\{tnhoefl,regrant,rbbrigh\}@sandia.gov}\newline\ 
\end{tabular}
}

\renewcommand{\shortauthors}{T. Hoefler et al.}

\begin{abstract}
Optimizing communication performance is imperative for large-scale
computing because communication overheads limit the strong
scalability of parallel applications.
Today's network cards contain rather powerful processors optimized for
data movement. However, these devices are limited to fixed functions,
such as remote direct memory access. 
We develop \abbr{}, a portable programming model to offload simple
packet processing functions to the network card. To demonstrate the
potential of the model, we design a cycle-accurate simulation
environment by combining the network simulator LogGOPSim and the CPU
simulator gem5.
We implement offloaded message matching, datatype processing, and
collective communications and demonstrate transparent full-application speedups. 
Furthermore, we show how \abbr{} can be used to accelerate
redundant in-memory filesystems and several other use cases. 
Our work investigates a portable packet-processing network acceleration
model similar to compute acceleration with CUDA or OpenCL. We show how
such network acceleration enables an eco-system that can significantly
speed up applications and system services. 
\end{abstract}

\begin{CCSXML}
<ccs2012>
<concept>
<concept_id>10010520.10010521.10010528.10010530</concept_id>
<concept_desc>Computer systems organization~Interconnection architectures</concept_desc>
<concept_significance>500</concept_significance>
</concept>
</ccs2012>
\end{CCSXML}

\maketitle

\begin{tikzpicture}[remember picture,overlay]
      \node[xshift=-1.5cm,yshift=-2.3cm] at (current page.north east)
      {\includegraphics[width=2cm]{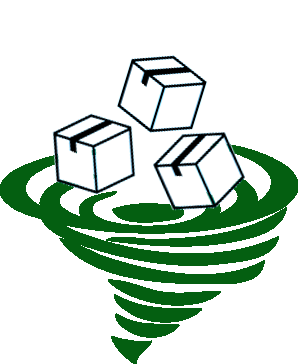}};
\end{tikzpicture}

\vspace{-2.8em}
\section{Motivation}

The current trend to move towards highly-scalable computing systems with
slow but energy-efficient processors increases the pressure on the
interconnection network. The recent leap in terms of bandwidth and
latency was achieved by removing the CPU from the packet processing
(data) path. Instead, specialized data processors offer remote direct
memory access (RDMA) functions and enable tens of gigabytes per second
transmission rates at sub-microsecond latencies in modern network
interface cards (NICs).

However, RDMA only transports data between (virtual) memories of
processes on different network endpoints. Different RDMA interfaces,
such as OFED~\cite{Hansen:2006:FRO:1188455.1188479}, uGNI/DMAPP~\cite{aries}, Portals 4~\cite{barrett2017portals}, or
FlexNIC~\cite{Kaufmann:2016:HPP:2954679.2872367} provide varying levels of
support for steering the data at the receiver. Yet, with upcoming
terabits-per-second networks~\cite{ethernet-roadmap}, we foresee a new
bottleneck when it comes to processing the delivered data: A modern CPU
requires 10-15ns to access L3 cache (Haswell: 34 cycles, Skylake: 44
cycles~\cite{Molka:2014:MMC:2618128.2618129,intel-manual}). However, a
400 Gib/s NIC can deliver a 64-Byte message each 1.2ns. 

The main problem is that packets are simply deposited into main
memory, irrespective of the contents of the message itself.
Many applications then analyze the received messages and rearrange them
into the application structures in host memory (e.g., halo exchanges,
parallel graph algorithms, database updates) even though this step can
logically be seen as part of the data routing.
This poses a barrier, very similar to pre-RDMA packet processing:
CPU cores are inefficient message processors because their
microarchitecture is optimized for computation. They require thread
activation, scheduling, and incoming data potentially pollutes the caches for
the main computation. 
Furthermore, due to the lack of a better interface, the highly-optimized
data-movement cores on the NIC are likely to place data
\emph{blindly} into host memory. 

To address these limitations and liberate NIC programming, we propose
\emph{streaming Processing in the Network} (\abbr{}), which aims to
extend the success of RDMA and receiver-based matching to simple
processing tasks that are dominated by data-movement.
In particular, we design a unified interface where programmers can
specify kernels, similar to
CUDA~\cite{Nickolls:2008:SPP:1365490.1365500} and
OpenCL~\cite{Stone:2010:OPP:622179.1803953}, that execute on the NIC. 
Differently from CUDA and OpenCL, kernels do not offload
compute-heavy tasks but data-movement-heavy tasks, specifically, tasks
that can be performed on incoming messages and only require limited local
state. Such tasks include starting communications with NIC-based
collectives, advanced data steering with MPI datatypes, data processing
such as network raid, compression, and database filters.
Similarly to OpenCL, \abbr{}'s interface is device- and
vendor-independent and can be implemented on a wide variety of systems.

\begin{figure}[h!]
  \vspace{-0.6em}
  \includegraphics[width=\columnwidth]{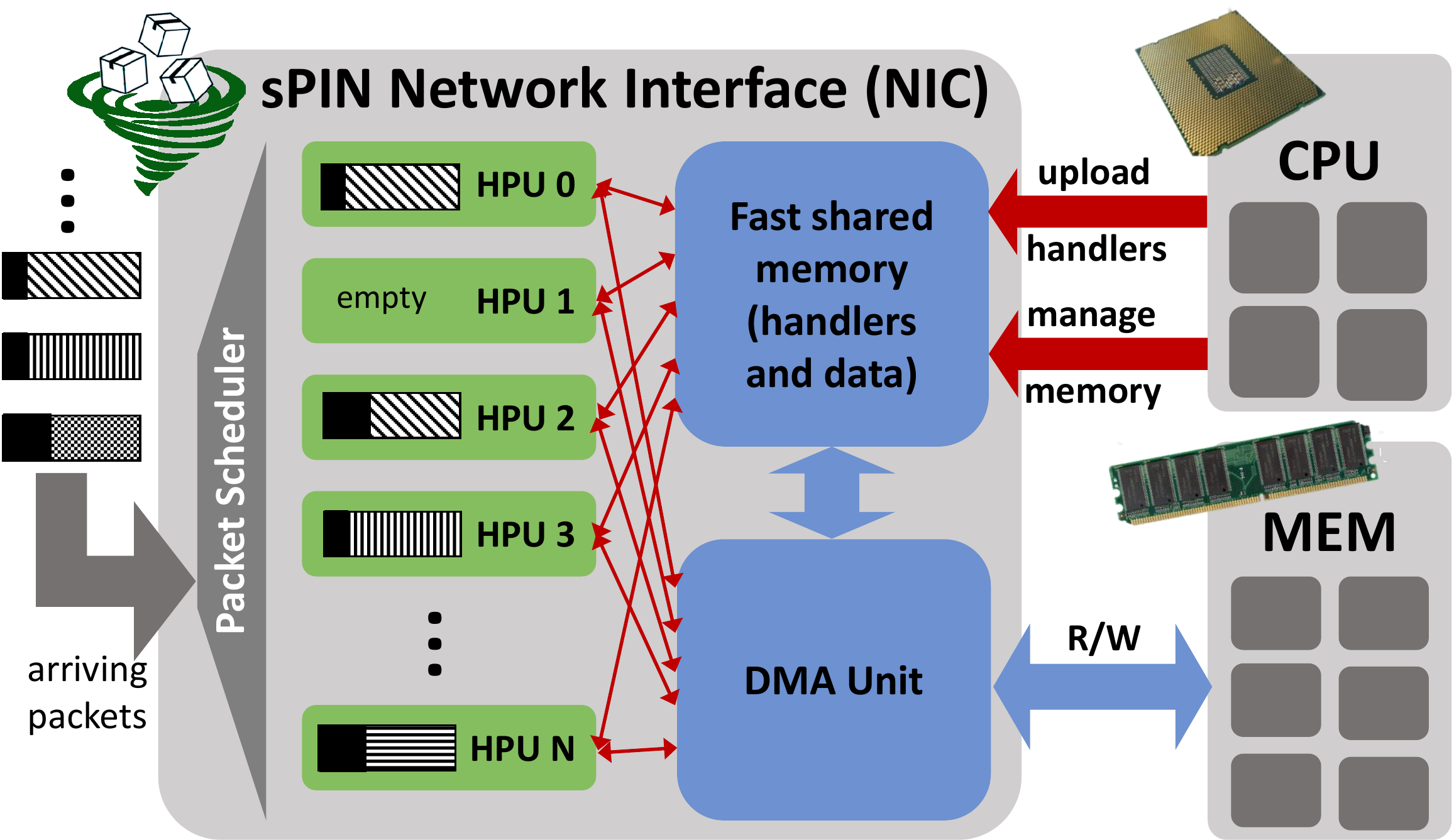}
  \vspace{-1.5em}
  \caption{\abbr{} Architecture Overview}
  \label{fig:overview}
  \centering
\end{figure}

We enable \abbr{} on existing NIC microarchitectures with typically
very small but fast memories without obstructing line-rate packet
processing. For this, we design \abbr{} around networking concepts such
as packetization, buffering, and packet steering. Packetization is the
most important concept in \abbr{} because unlike other networking layers
that operate on the basis of messages, \abbr{} exposes packetization to
the programmer.
Programmers define \emph{header}, \emph{payload}, and \emph{completion
handlers} (kernels) that are executed in a \emph{streaming way} by
handler processing units (HPUs) for the respective packets of each
matching message. Handlers can access packets in fast local memory
and they can communicate through shared memory.  sPIN offers protection
and isolation for user-level applications and can thus be implemented in
any environment. Figure~\ref{fig:overview} shows sPIN's architecture. 
\vspace{-1em}

\abbr{}'s philosophy is to expose the highly-specialized packet
processors in modern NICs to process short user-defined functions. By
``short'', we mean not more than a few hundred instructions from a very
simple instruction set. In this sense, handlers are essentially
pluggable components of the NIC firmware. \abbr{} offers unprecedented
opportunities to accelerate network protocols and simple packet
processing functions and it can be implemented in discrete NICs, SoCs,
or even in parts of CPU cores. Offering programmable network devices
liberates the programmer from restricted firmware functionalities and
custom accelerators and is
the next step towards full software-defined networking infrastructures.
The following C code demonstrates how to define handler functions in a
user-application:
\begin{lstlisting}[basicstyle=\ttfamily\footnotesize]
__handler int header_handler(const ptl_header_t h, void  *state) {
  /* header handler code */ }

__handler int payload_handler(const ptl_payload_t p, void *state) {
  /* packet content handler code */ }

__handler int completion_handler(int dropped_bytes, bool flow_control_triggered, void *state) {
  /* post-message handler code */ }

channel_id_t connect( peer, /* ... */, &header_handler, 
               &payload_handler, &completion_handler );
\end{lstlisting}
The function decoration \lstinline$__handler$ indicates that this
function must be compiled for the sPIN device. Handler code
is passed at connection establishment. 
This allows a single process to install
different handlers for different connections. Arguments are the packet
data and \lstinline$*state$, which references local memory that is shared
among handlers.

As a principled approach to network offloading, \abbr{} has the
potential to replace specific offload solutions such as ConnectX CORE-Direct
collective offload~\cite{6008920}, Cray Aries~\cite{aries}, IBM
PERCS~\cite{ibm-percs-network}, or Portals 4~\cite{hemmert2010using}
triggered operations. Instead, the community can focus on developing
domain or application-specific \abbr{} libraries to accelerate
networking, very much like NVIDIA's cuBLAS or
ViennaCL~\cite{Rupp:ViennaCL}. A vendor-independent interface would
enable a strong collaborative open-source environment similar to the
Message Passing Interface (MPI) while vendors can still distinguish
themselves by
the design of NICs (e.g., specialized architectures for packet
processing such as massive multithreading in Intel's Network Flow
Processor).

Specifically, in this paper, we 
\begin{itemize}[noitemsep,topsep=2pt,leftmargin=20pt]
 \item present the design of an acceleration system for NIC offload;
 \item outline a microarchitecture for offload-enabled smart NICs;
 \item design a cycle-accurate validated simulation environment integrating
   network, offload-enabled NICs, and CPUs;
 \item outline and analyze use cases for parallel applications as well
   as for distributed data management systems;
 \item and demonstrate speedups for various real applications.
\end{itemize}

\subsection{Background}\label{sec:background}

\hpara{SMP vs. AM}
We now provide a brief overview of related technologies. At first
glance, \abbr{} may seem similar to active messages
(AM)~\cite{vonEicken:1992:AMM:146628.140382}---it certainly
shares many potential use cases. Yet, it is very different because it
specifies an architecture for fast and tightly integrated NIC packet
processing.
Both, AM and \abbr{} are independent of process scheduling at the host
OS and can be defined independently of the target hardware.
The major difference is that AMs are invoked on full messages while
\abbr{} is defined in a streaming manner on a per-packet basis. Early AM
systems that constrained the message size
may be considered as special cases of \abbr{}. Yet, \abbr{} enables to
pipeline packet processing, similarly to wormhole routing while AM would
correspond to store and forward routing. 
Furthermore, AMs use host memory for buffering messages while \abbr{}
stores packets in fast buffer memory on the NIC close to the processing
units for fastest access; accesses to host memory are possible but should
be minimized. 
A last semantic difference is that in AM, a message can be considered
atomic because a handler is invoked after the message is delivered while
in \abbr{} handlers are invoked on parts of a message and only those
parts (i.e., packets) can be processed atomically. 

\hpara{SMP vs. packet processing in hardware}
\abbr{} is in fact closer to packet processing systems than to AM. Fast
packet processing hardware has been designed in the Intel IXP family of
chips and continued in the Netronome NFP series (cf.~\cite{gavrilovska}). Recent progress in
software defined networking (SDN) enables users to program switches with
simple parse-match-action rules that allow simple packet processing and
routing in the network. P4~\cite{bosshart2014p4} is a language to express such
rules concisely and it supports fast packet parsing in hardware. Another
related proposal is
FlexNIC~\cite{Kaufmann:2016:HPP:2954679.2872367}, which builds on the
SDN/P4 ideas and extends routing to the DMA engine in the NIC. Yet, in
the HPC context, this routing is comparable to what current HPC network
interfaces such as Portals 4 already support in hardware (receiver-side
steering). \abbr{} goes far beyond these to exploit processing of packets on
specialized units in fast local memories.

\section{Processing in the Network}

\hpara{messages vs. packets}
\abbr{}'s central philosophy, which is independent of any particular
implementation, is based on the fact that network devices split messages
into packets for transmission (messages correspond to network transactions). Packets are easier to manage because they
can be buffered and forwarded independently. We adopt this philosophy
for \abbr{} to enable a direct implementation in a network device. In
\abbr{}, the programmer defines handler functions that execute on a set
of packets that logically form a message. Those functions are executed
on one or multiple handler processing units (HPUs). A simple runtime
system is responsible for controlling the handlers and scheduling them
for execution on HPUs. Each handler owns shared memory that is
persistent across the lifetime of a message, i.e., handlers can use that
memory to communicate. 

\begin{figure}[h!]
  \vspace{-0.7em}
  \includegraphics[width=\columnwidth]{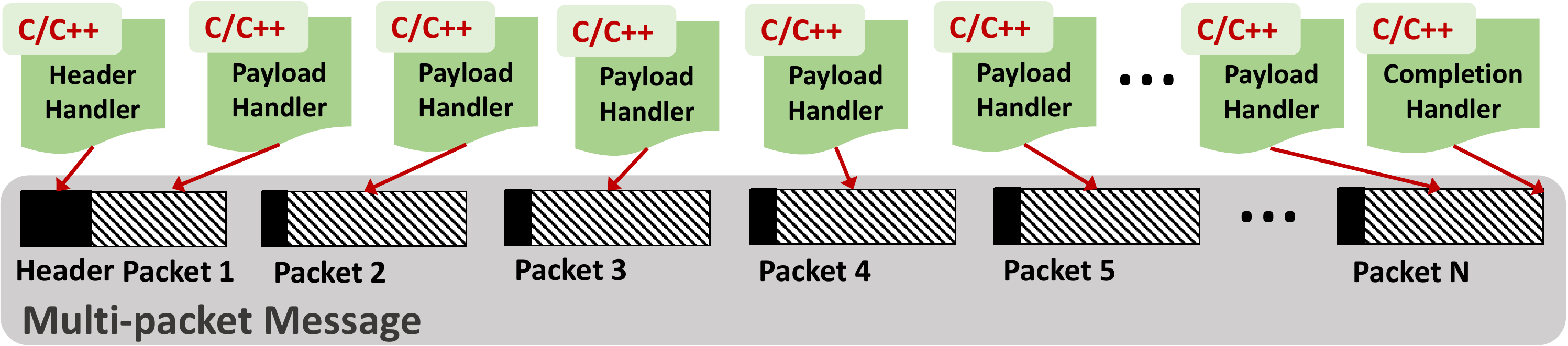}
  \vspace{-2.0em}
  \caption{\abbr{} Message Handler Overview}
  \vspace{-1.0em}
  \label{fig:messagehandlers}
  \centering
\end{figure}
\hpara{handler types}
Figure~\ref{fig:messagehandlers} shows how the handlers relate to parts
of the message.
Network layers enforce that all necessary information to identify a
message and steer it into memory is included in the first packet that
we call \emph{header packet}. 
Many communication systems, such as Ethernet, replicate this header
information in each packet (black boxes). 
To enable fast channel-based systems, \abbr{} does not rely on
replicated header information but delegates to the NIC-based runtime system to
identify the set of packets that belongs to the same message. 
\abbr{} defines three handler types to be invoked on different parts of
a message: the header handler works on header information, the payload
handler processes the message payload after the header handler completes,
and the completion handler executes after all instances of a payload
handler have completed. There is no requirement that packets arrive
or are processed in order.

\hpara{resource management}
HPU memory is managed by the host operating system and the user-level application using
a typical control-/data-plane separation.  Performance-critical parts
such as invoking handlers are performed without OS involvement while
other parts, such as isolation, setting up protection domains, and
memory management on the device can be performed through the host OS.
The host compiles and offloads handler code to the HPU, similar to how GPU
kernels work. Handlers only contain a few hundred instructions and their code
can thus be placed into fast memory for quick execution. The system can
reject handler code that is too large. The application on the CPU allocates and
initializes HPU memory at handler-installation time. Handlers can
communicate through shared memory but they cannot perform any memory
management and their local memory only offers linear (physical)
addressing. 

\hpara{programming model}
Handlers are programmed by the user as standard C/C++ code to enable
portable execution and convenient programming. They can only contain
plain code, no system calls or complex libraries.  Handlers are then
compiled to the specific target network ISA. The program can contain
static segments of pre-initialized data.  Handlers are not limited in
their execution time, yet, resources are accounted for on a per-application
basis. This means that if handlers consume too much time, they may stall
the NIC or drop packets. Thus, programmers should ensure that
handlers can operate at line-rate on the target device. It is key that
handlers can be started at very low cost for each packet; we assume that
execution can start within a cycle after a packet arrived in the buffer
(assuming a HPU is available).  Furthermore, to guarantee high
message rate, handlers need to be set up quickly from the host and
parameters must be passed with low overhead. Handlers execute in a sandbox with
respect to application memory, i.e., they may only access a restricted
memory range in the application's virtual address space.

\enlargethispage{2em}
\hpara{local handler actions}
Handlers can perform various actions besides executing normal C code.
Ideally, these actions are implemented as hardware instructions.  At the
start of a handler, the packet is available in a fast buffer (ideally
single-cycle access). Handlers have access to host memory via DMA.
This enables the runtime system to deschedule handlers from
massively-threaded HPUs while they are waiting for host memory. Handlers
do not block and 
can voluntarily yield to another handler. Yet, it is a central part
of the programming philosophy that DMA should be used scarcely, as it is
expensive and its performance is non-deterministic. 
Handlers can generate two types of messages: (1) messages originating
from HPU memory and (2) messages originating from host memory.  Messages
issued from HPU memory can only contain a single packet and are thus
limited to the MTU. Messages issued from HPU memory may block the HPU
thread until the message is delivered (i.e., the NIC may use HPU memory as
outgoing buffer space). Messages issued from host memory shall enter the
normal send queue as if they were initiated from the host itself.
Messages from host memory shall be nonblocking.

\section{A complete \abbr{} Interface}

\abbr{} can be added to any RDMA network. As an example to demonstrate
all \abbr{} features, we use the Portals 4 network interface because it
offers advanced receiver-side steering (matching), OS bypass,
protection, and NIC resource management. It has been implemented in
hardware and demonstrated to deliver line-rate interactions with the
host CPU~\cite{bxi}. Furthermore, its specification is openly available; we briefly
summarize the key aspects in the following.

\subsection{Overview of Portals 4}
\hpara{virtual NIs}
Portals 4 specifies logical and physical addressing modes and
offers matched or unmatched operation for logical network interfaces
that are bound to physical network resources.
Logical addressing can be used by runtime systems to offer a NIC
accelerated virtualized process space.
A matched interface allows the user to specify match bits to direct
incoming messages to different logical lists identified by tags (potentially
with a mask) in a single match list. Each logical queue head specifies a
steering action for incoming packets.
Without loss of generality, we focus on logically addressed and matched
mode as this combination provides the highest level of
abstraction~\cite{barrett2017portals}.

\hpara{communication and completion}
Portals 4 offers put, get, and atomic communication operations.
Completion notification occurs through counting events or appending
a full event to an event queue, which is also used for error notification.
Memory descriptors (MDs) form an abstraction of memory to be sent;
counters and event queues are attached to it. Matching entries (MEs)
identify receive memory; matching is performed through a 64-bit masked
id; MEs have counters and event queues associated with them. Portals 4 offers full
memory access protection through MDs and MEs.

\hpara{SMP extends Portals 4 and integrates into it}
\sloppy
Portals 4 offers two mechanisms that go beyond simple message steering
and that allow implementation of a limited Network Instruction Set
Architecture (NISA)~\cite{hemmert2010using,exploiting-offload-enabled-ni}. First,
it enables communications that have been previously set up
to be triggered by counters reaching a certain threshold. Second, Portals 4 MEs can
have locally managed offsets where incoming data is efficiently packed
into buffers using a local index. Both mechanisms are limited because only incoming
messages can trigger and steer operations, not the data in those
messages. These actions also cannot process packets (local atomics can
be used to emulate very limited processing capabilities~\cite{exploiting-offload-enabled-ni}).
\abbr{}
integrates with and extends Portals 4 to offer more powerful message
steering, protocol implementation, and packet processing
functionalities.

\subsection{A sPIN Interface for Portals 4}

\hpara{semantics}
Based on the general semantics for \abbr{} systems, we now derive a
candidate Portals 4 (P4) interface called P4sPIN. We only provide an
overview of the functions in the main part of the paper and refer to the
appendix for signatures and a detailed description.
All packet handlers
are associated with a specific matching entry (ME) to which incoming
messages are matched. MEs are posted using 
\lstinline$PtlMEAppend$ (cf.~Appendix~\ref{sec:p4md}).  We extend
this call to enable registering handlers with 
additional arguments to identify the three handlers, the shared memory
setup, the initial memory state to initialize the shared memory, and the
handler's memory region at the host (if needed). 

\hpara{HPU memory}
The ME requires a handle that identifies an HPU shared memory space to
run the handler in. HPU memory is allocated using the 
\lstinline$PtlHPUAllocMem$ function (see~Appendix~\ref{sec:hpumem}) at
the host (before handler installation).
This explicit management allows the user to re-use the same HPU memory
for multiple MEs. HPU memory remains valid until it is deallocated. If
multiple incoming messages match MEs that specify the same HPU memory
then the handlers should perform concurrency control.

\hpara{flow control}
If an incoming packet arrives and matches an ME but no HPU execution
contexts are available, the NIC may trigger flow control for the
respective portal table entry. This is symmetric to the situation where
the host runs out of compute resources and fails to post new MEs to the
NIC. In a flow control situation, packets arriving at a specific portal
table entry are dropped until the entry is re-enabled. Note that this
can happen during the processing of a message. In this case, the
completion handler is invoked and notified through the flag
\lstinline$flow_control_triggered$.

\subsubsection{Header Handler}

The header handler is called exactly once per message and no other
handler for this message is started before the header handler completes.
It has access to only the header fields that can include user-defined
headers (the first bytes of the payload). User-defined headers are
declared statically in a struct to enable fast parsing in hardware. Yet,
the struct offers flexibility as it guarantees no specific memory
layout. For example, pre-defined headers could be in special registers
while user-defined headers can reside in HPU memory. The struct is
static such that it can only be used on the right-hand side of
expressions. This makes it possible to implement using registers.

\subsubsection{Payload Handler}

The payload handler is called after the header handler completes for 
packets carrying a  payload.
The passed payload
does not include the part of the user-header that was specified by
\lstinline$user_hdr$.

Multiple instances of the payload handler can be executed in parallel and the programmer
must account for concurrent execution. Payload handlers share all HPU
memory coherently. The illusion of private memory
can be created by the programmer, yet no protection exists. To create
private memory, the system offers a compile-time constant
\lstinline$PTL_NUM_HPUS$ that contains the number of handle execution units.
Note that each unit may be used to
process multiple payload handlers serially but only
\lstinline$PTL_NUM_HPUS$ handlers can be active simultaneously at any
given time. Furthermore, a runtime constant \lstinline$PTL_MY_HPU$
allows a running handler to determine on which HPU it is running.
Handlers may not migrate between HPUs while they are running. These two
constants allow the user to allocate arrays of size
\lstinline$PTL_NUM_HPUS$ and index into them using
\lstinline$PTL_MY_HPU$ to emulate HPU-private data.

\subsubsection{Completion Handler}

The completion handler is called once per message after all header handlers
and payload handlers have completed but before the completion event is delivered to the
host. The handler can be used for final data collection or cleanup tasks
after the message has been processed. 
The value in \lstinline$dropped_bytes$ indicates how many bytes of
payload data have been dropped by payload handlers. Bytes can either be
dropped by payload handlers returning a variant of \lstinline$DROP$ or
if a flow-control situation occurs. The flag
\lstinline$flow_control_triggered$ indicates that flow control was
triggered during the processing of this message and thus some packets
may have been dropped without being processed by payload handlers.  
The pointer state points at the initial data in HPU memory. This data
may have been initialized by the host or header handler.

All handlers can perform various actions as described before. 
The detailed interfaces for all calls are specified in
Appendix~\ref{sec:hdlract}.

\section{Prototyping \abbr{}}

We now describe two prototype implementations of \abbr{} as an NISA. The first architecture represents a
\emph{discrete} network card (``dis'') that is attached to the CPU via a
chip-to-chip interconnect such as PCI express (PCIe). The second
architecture represents an \emph{integrated} network card (``int'') that is on the
same chip as the CPU cores and attached via a fast signaling protocol
such as the Advanced eXtensible Interface (AXI). 

\subsection{HPU Design}

\hpara{intro}
The HPU architecture is an integral part of the \abbr{} design. We
briefly describe some design, optimization, and customization ideas
without proposing any particular architecture. We assume that most of
today's NIC architectures can be re-used. \abbr{} can be conceptualized ass
being equivalent to installing custom mini-firmware for each application on the NIC. 

\hpara{HPU}
The header processing unit (HPU) should have fastest (ideally single-cycle)
access to local memory and packet buffers. To achieve this, it could be
connected to packet buffers directly. HPU memory is not cached. Most HPU
instructions should be executed in a single cycle and the documentation
should be explicit about instruction costs. Handlers should be invoked
immediately after a packet arrives or the previous handler completes.
Handlers require no initialization, loading, or other boot activities
because all their context is pre-loaded and memory is pre-initialized.
HPUs can be implemented using massive multithreading to utilize the
execution units most efficiently. For example, if handler threads wait
for DMA accesses, they could be descheduled to make room for different
threads. Only the context would need to be stored, similarly to GPU
architectures.

\hpara{needed memory overhead}
Handler execution will delay each message that is processed. This requires
enough HPU cores to process at full bandwidth and additional
memory. The required memory overhead can be computed using Little's law.
If we assume a handler executes between 10 and 500 instructions at 2.5GHz
and IPC=1, we expect a maximum delay of 200ns per packet. With 1Tb/s, we
can calculate the overhead as 1 Tb/s $\cdot$ 200ns = 25 kB. We expect that
this can easily be made available and more space can be added to hide
more latency, probably up to several microseconds.

\hpara{can be implemented anywhere}
\abbr{} can be implemented in multiple different environments.  On a
discrete NIC, one can take advantage of the existing packet processing
infrastructure and buffer space. \abbr{} can also be added to an SoC to
steer messages to the correct cores for processing. At the other
extreme, \abbr{} can be implemented in a core with an integrated NIC as
a small layer between the transceiver and the core. It could even use the
pipeline of a super-scalar core with tagged instructions. 

\subsection{Simulation Setup}

To evaluate \abbr{} at scale, we combine two open-source simulators that
have been vetted extensively by the research community:
LogGOPSim~\cite{hoefler-loggopsim} to simulate the network of parallel
distributed memory applications and
gem5~\cite{Binkert:2011:GS:2024716.2024718} to simulate various CPU and
HPU configurations.  LogGOPSim supports the simulation of MPI
applications, injection of system
noise~\cite{hoefler-noise-sim,hoefler-collnetnoise}, and has been
calibrated and validated on InfiniBand
clusters~\cite{hoefler-loggopsim}.  The cycle-accurate gem5 simulator
supports a large number of different CPU configurations and is thus an
ideal choice for our designs. 

In our setup, LogGOPSim is driving the simulation by running a
trace-based discrete-event loop. Traced events are all Portals 4
and MPI functions as well as the invocation of handlers. LogGOPSim invokes
gem5 for each handler execution and measures the execution time. The two
simulators communicate via a special memory-mapped region in gem5
through which an executing handler can issue \emph{simcalls} from gem5 to
LogGOPSim. Simcalls enable simulated applications in gem5 to invoke
functionality in LogGOPSim, for example, to insert new messages into the
discrete-event queue.  Overall, this combination of trace-based network
simulation and cycle-accurate execution-based CPU simulation enables
high-accuracy and efficient modeling of the complete \abbr{} system. 

We parametrize for a future InfiniBand system using the LogP model extended with
a packet-level simulation to model the $L$ (Latency) parameter more accurately.
The injection overhead is not parallelizable, thus, we use $o=65ns$
(injection overhead),
which we measured on a modern InfiniBand system. Similarly, we expect
the message rate to stay approximately similar, around 150 Million
messages per second for Mellanox ConnectX-4~\cite{mlx-talk} and thus set
$g=6.7ns$ (inter-message gap). As bandwidth, we expect networks to deliver 400 Gib/s around
the deployment time of~\abbr{} and thus set $G=2.5ps$ (inter-Byte gap). The latency is
determined by a model for a packet-switched network where we assume a
switch traversal time of $50ns$ (as measured on modern switches) and a
wire length of $10m$ (delay of $33.4ns$). We construct a fat tree
network from 36-port switches. The model is simulated using
the LogGOPSim MPI simulator that has been shown to be within 10\%
accuracy when compared with real runs~\cite{hoefler-noise-sim}. 

We model each NIC to have four 2.5GHz ARM Cortex A15 out-of-order HPU cores
using the ARMv8-A 32-bit ISA. We configure the cores without cache and
with gem5's SimpleMemory module configured as scratchpad memory that can
be accessed in $k$ cycles (we use $k=1$ in the paper). Endo et
al.~\cite{gem5-microarch} demonstrated that the average absolute error
of a comparable ARM Cortex A15 was only 7\% when compared to real hardware.
Messages are matched in hardware and only header packets search the
full matching queue. A matched header packet will install a channel into
a fast content-addressable memory (CAM) for the remaining packets. We assume
that matching a header packet takes 30 ns
(cf.~\cite{Underwood:2011:EFC:2057181.2057595}) and each
following packet takes 2ns for the CAM lookup. We assume that matching
and the network gap ($g$) can proceed in parallel.

We model the host CPU as eight 2.5Ghz Intel Haswell cores 
with 8 MiB cache and a DRAM latency and bandwidth of 51 ns and 150
GiB/s, respectively. A similar configuration has been analyzed by Akram
et al.~\cite{akram201686}.

\subsection{DMA and Memory Contention}

HPU accesses to host memory are performed via DMA. We extended the
simulator by adding support to model contention for host memory.
This contention either happens through the north-bridge via
PCIe (discrete NIC) or through the memory controller of the CPU
(integrated NIC). We model DMA at each host as a simple LogGP
system~\cite{pciemodel,Martinasso:2016:PCP:3014904.3014989}. We set
$o=0$ and $g=0$ because these times are already captured by the
cycle-accurate gem5 simulation when initiating the request. We set $L$
and $G$ depending on the discrete or integrated HPU
configuration as follows.

The discrete NIC is connected through an off-chip interconnect such as
PCI express. We use 32-lane PCI express version 4 as a candidate
system with a latency of $L=250ns$, and $G=15.6ps$ (64 GiB/s).
The integrated NIC is connected directly to the chip's memory
controller, which allows a much lower latency of $L=50ns$ and the same
bandwidth as the main CPU $G=6.7ps$ (150 GiB/s).

The DMA time is added to the message transmission when the NIC delivers
data into host memory (e.g., for every message in RDMA and Portals 4),
for HPU calls PutFromHost, and when the HPU invokes DMA routines to main
memory.

\subsection{Microbenchmarks}

We first demonstrate the parameters of \abbr{} with a set of
microbenchmarks before we show a series of use cases for
real-world applications. 

\begin{figure*}[ht!]%
    \centering
    \subfloat[Schematic Ping Pong (thick lines show multi-packet messages,
    DMA transactions at source omitted for clarity)]{
  \includegraphics[height=3.72cm]{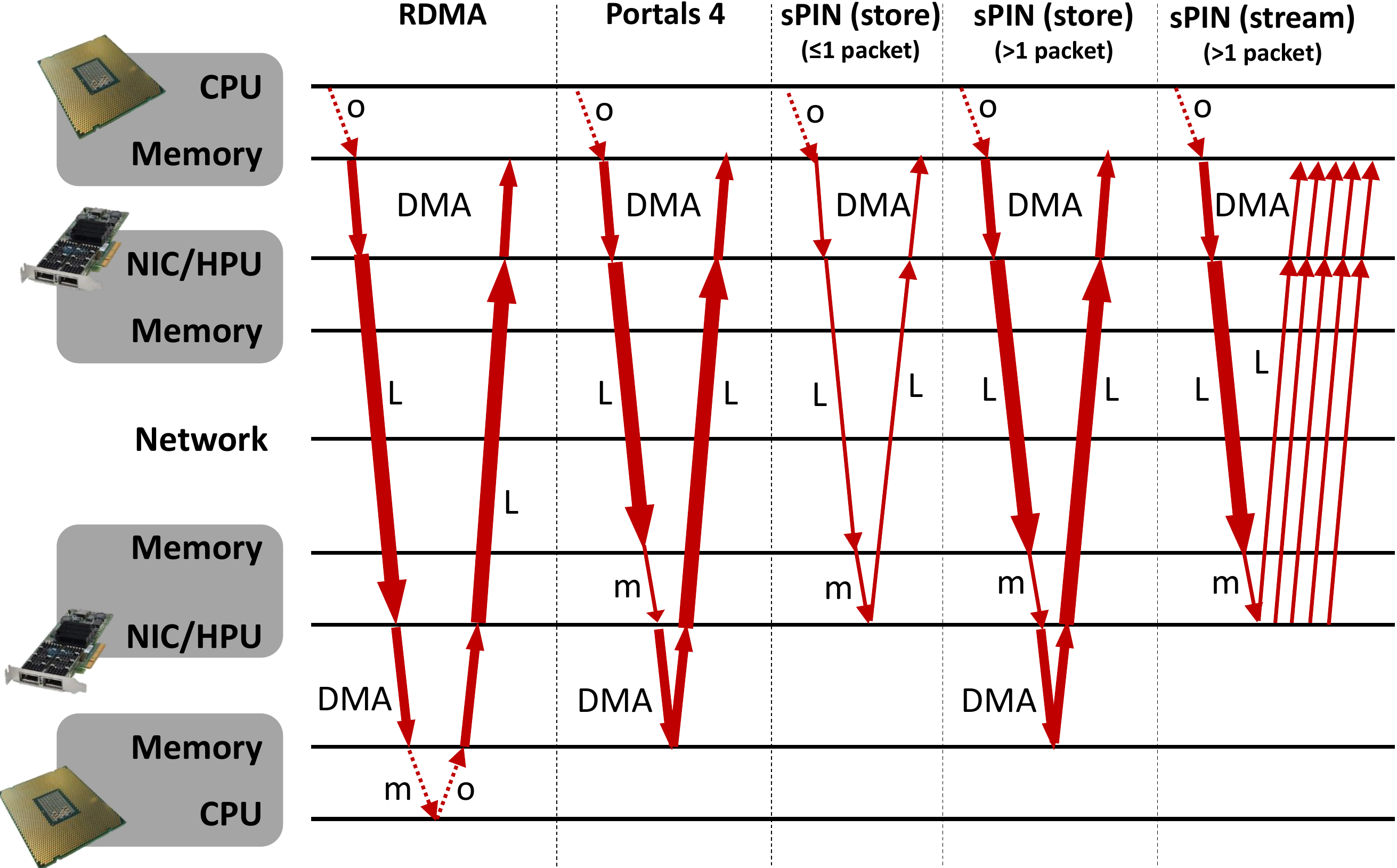} \label{fig:pingpongdiagram}
}%
    \hfill
    \subfloat[Ping Pong (integrated NIC)]{{
	\includegraphics[height=3.72cm]{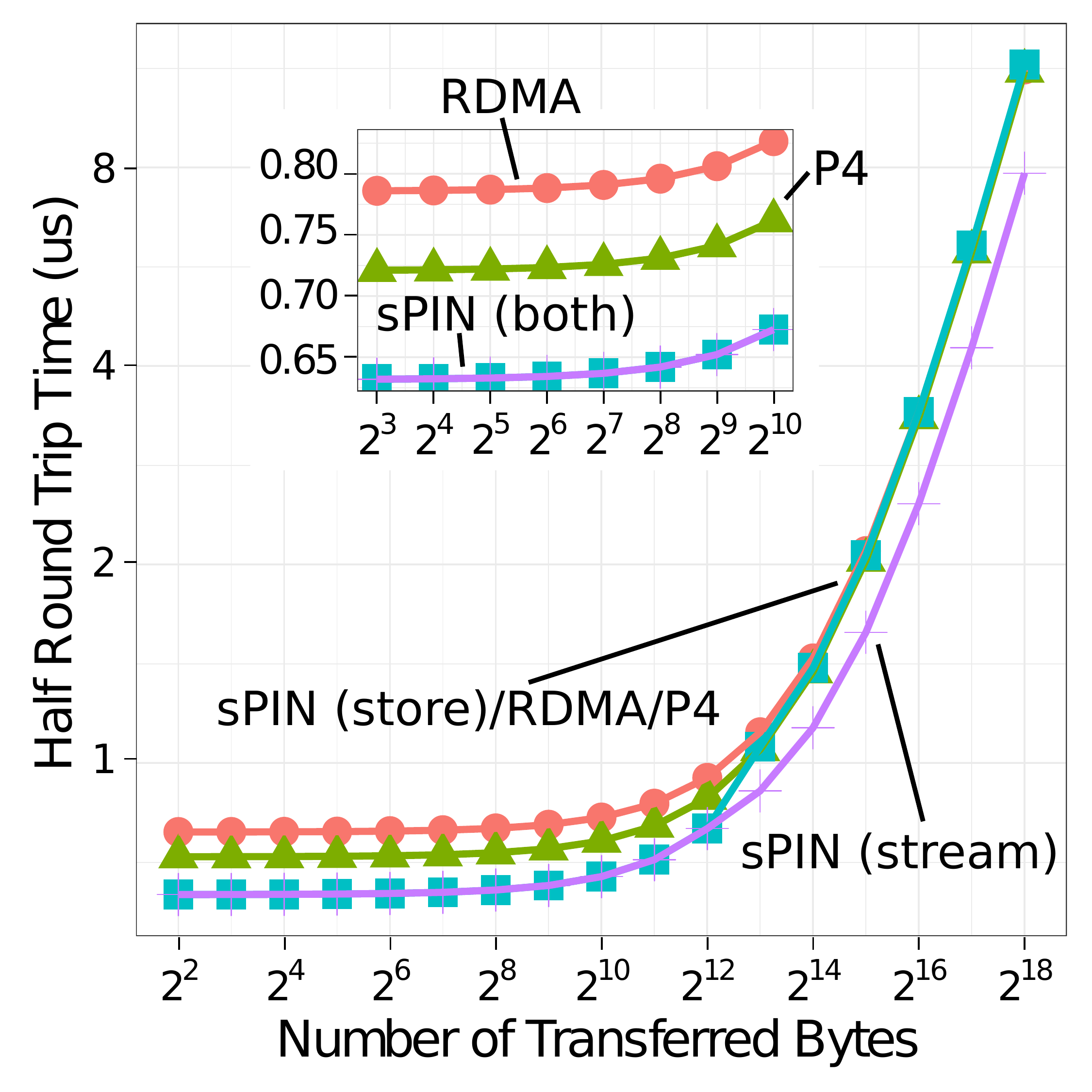}
  \label{fig:pingpongintegrated}
}}%
    \hfill
    \subfloat[Ping Pong (discrete NIC)]{{
	\includegraphics[height=3.72cm]{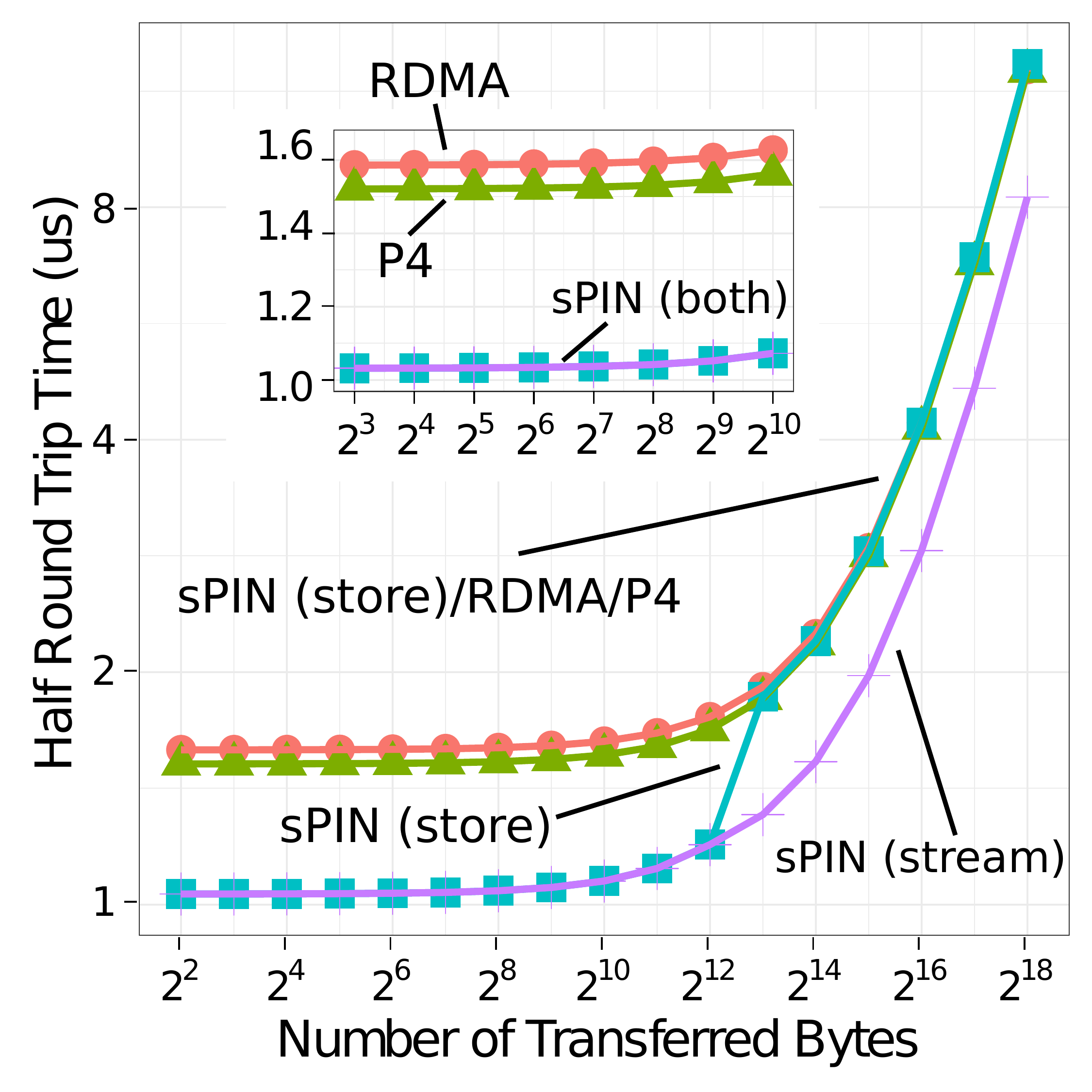}
  \label{fig:pingpongdiscrete}
}}%
    \hfill
    \subfloat[Accumulate (both types)]{{
	\includegraphics[height=3.72cm]{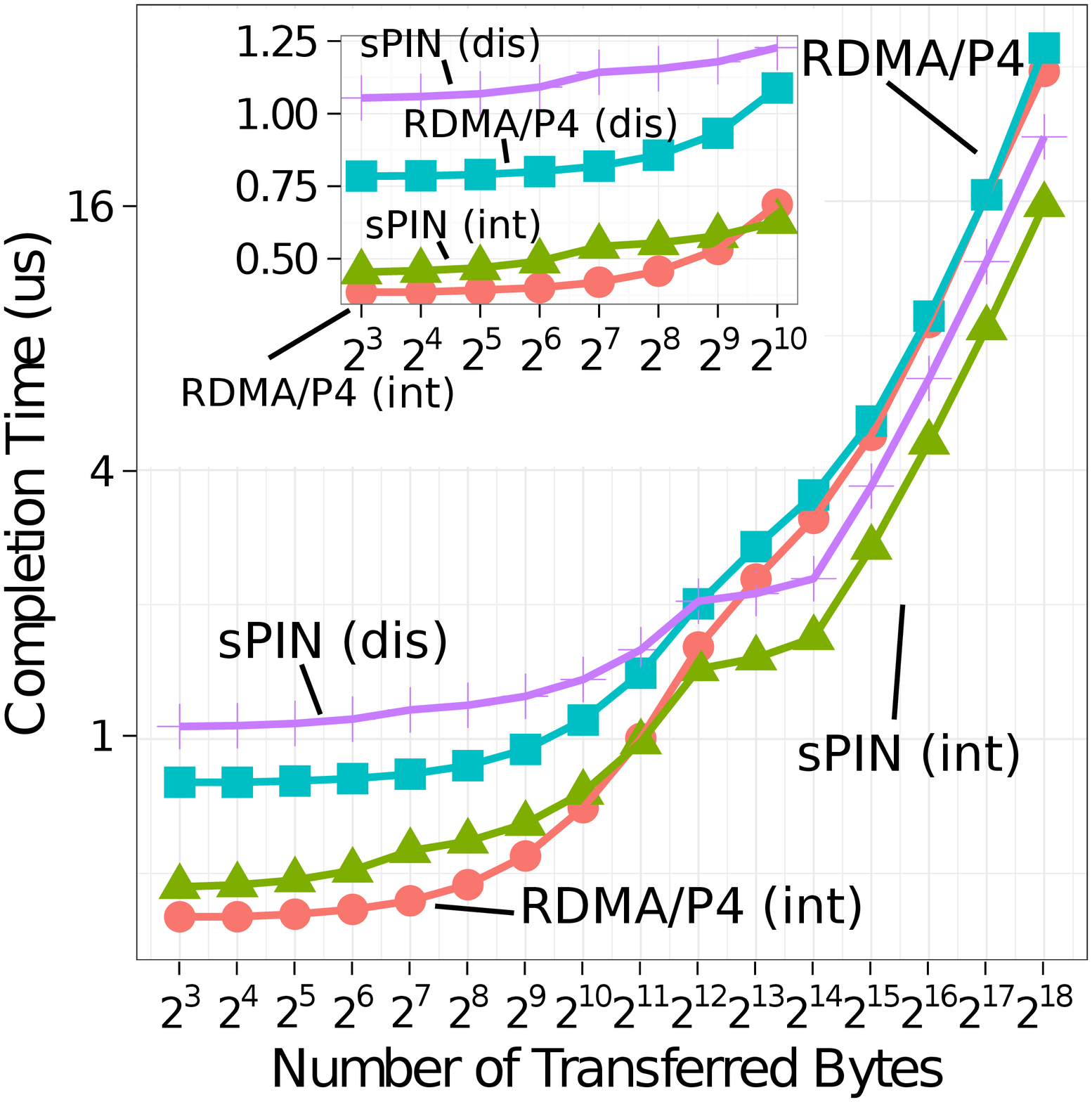}
  \label{fig:accumulate}
}}%
  \vspace{-1em}
    \caption{Ping pong and remote accumulate comparing RDMA, Portals 4, and various \abbr{} implementations}%
  \vspace{-1em}
\end{figure*}

\subsubsection{Ping-Pong Latency}

We compare our two \abbr{} systems with standard RDMA as well as 
Portals 4 with a simple ping-pong benchmark. This illustrates 
the basic capabilities of processing messages on the NIC. For RDMA
and Portals 4, all messages need to be stored to and loaded from main
memory. \abbr{} can avoid this memory traffic and reply directly from
the NIC buffer, leading to a lower latency and less memory traffic at the
host.  
Figure~\ref{fig:pingpongdiagram} illustrates the
following explanations of time spent on the CPU, the host memory, the
NIC, and its memory when executing ping-pong. All variants start a
transmission from the main CPU and the message travels for time $L$
through the network.

For RDMA, the pong message is sent by the main CPU. Thus, the
destination CPU polls for a completion entry of the incoming ping
message, performs message matching, and immediately posts the pong
message. The completion will only appear after the whole message has
been deposited into host memory. Processing occurs on the CPU,
therefore, system noise may delay the operation.   

For Portals 4, the pong message is pre-set up by the destination CPU and
the reply is automatically triggered after the incoming message has been
deposited into host memory. Thus, system noise on the main CPU will not
influence the pong message. Even though the message itself is
automatically triggered, the data is fetched via DMA from the CPU's main
memory as in the RDMA case. 

In \abbr{} ping-pong, the ping message may invoke header, payload,
and/or completion handlers. 
\abbr{} gives us
multiple options for generating the pong message: (1) (store) the ping
message consists of a single packet and a pong can be issued with a put
from device, (2) (store) the ping message is larger than a packet and
the pong message is issued with put from host using the completion
handler after the packet is delivered to host memory, and (3) (stream) a
payload handler could generate a pong put from device for each incoming
packet. Here the NIC would act as a packet processor and split a
multi-packet message into many single-packet messages. The first two
correspond to store and forward processing for different message sizes
while the last corresponds to fully streaming operation.

The performance of ping-pong for all configurations is shown for
integrated \abbr{} implementations in
Figure~\ref{fig:pingpongintegrated} and for discrete implementations in
Figure~\ref{fig:pingpongdiscrete}. The latency difference is more
pronounced in the discrete setting due to the higher DMA latency. Large
messages benefit in both settings from the streaming approach where data
is never committed to the host memory. The full handler code is shown in
Appendix~\ref{sec:pingpongcode}.

\subsubsection{\abbr{} Accumulate}

In the second microbenchmark we evaluate \abbr{}'s interaction with
local memory at the destination. For this, we choose a simple accumulate
benchmark where we send an array of double complex numbers to be
multiplied to an array of numbers of the same type at the destination.
The multiplication is either performed on the CPU or by the NIC/HPU.
This example represents an operation that is not typically supported as
a NIC atomic in RDMA or Portals 4 NICs. Yet, it can easily be
implemented using \abbr{}. If the operation was supported by the NIC
directly, then the performance would be similar to \abbr{}.

In an RDMA implementation, the data would be delivered into a temporary
buffer that is read by the CPU and then accumulated into the
destination buffer. Here, the NIC writes the array to host memory,
notifies the CPU, which then reads two arrays from host memory and
writes the result back. So if the data is of size $N$, we have two $N$-sized read
and two $N$-sized write transactions.

In \abbr{}, the packets will arrive and be scheduled to different HPUs.
Each HPU will fetch the data from host memory, apply the operation, and
write it back. For an array of size $N$, we only read $N$ bytes and
write $N$ bytes. Thus, \abbr{} halves the memory load compared to RDMA
and P4.
However, because the data has to be moved twice through the bus from the
host memory, the bus latency may slow-down processing of small messages.
Many NICs employ caching of atomic operations to hide the DMA
latency~\cite{aries} by relaxing the memory coherence---\abbr{} can
use similar techniques but we decided to show a coherent system with the
latency overhead.

Figure~\ref{fig:accumulate} shows the accumulate results. As expected,
the latency for small accumulates is higher for \abbr{} than for RDMA
because the data has to be first fetched via DMA to the HPU. This is
especially pronounced for the discrete NIC configuration where we see
the 250ns DMA latency. However, due to \abbr{}'s streaming parallelism
and the resulting pipelining of DMA requests, processing large
accumulates gets significantly faster for larger messages. The full handler
code is shown in Appendix~\ref{sec:accumulatecode}.

\paragraph{How many HPUs are needed?} We use this example to discuss one of the most important design choices
of a \abbr{} NIC: the number of HPU cores. Each packet is processed by an
HPU and multiple packets belonging the same or different messages may be
processed in parallel. Now we can discuss the number of needed HPUs in
order to guarantee line-rate execution. This can be modeled by Little's
law. If we assume an average execution time per packet of
$\overline{T}$ and an expected arrival rate of $\overline{\Delta}$, then
we need $\overline{T}\cdot \overline{\Delta}$ HPUs in the system. With a
fixed bandwidth ($1/G$), the arrival rate only depends on the
packet size $s$ and the gap $g$ such that $\overline{\Delta}=\min\{1/g,
1/(G\cdot s)\}$. 
\begin{figure}[h!]
  \vspace{-0.6em}
  \includegraphics[width=\columnwidth]{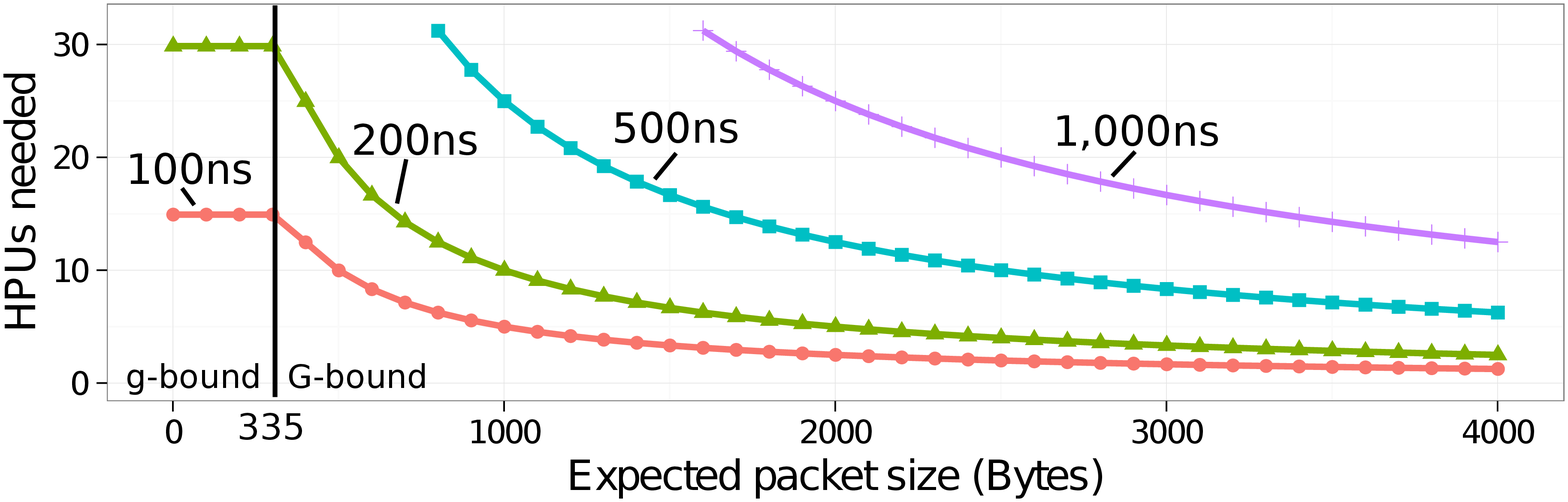}
  \vspace{-2.3em}
  \caption{HPUs needed depending on $\overline{T}$ and $s$.}
  \vspace{-1.0em}
  \label{fig:hpusneeded}
  \centering
\end{figure}
\begin{figure*}[t]%
    \centering
    \subfloat[Broadcast on a binomial tree (discrete NIC)]{
  \includegraphics[height=3.82cm]{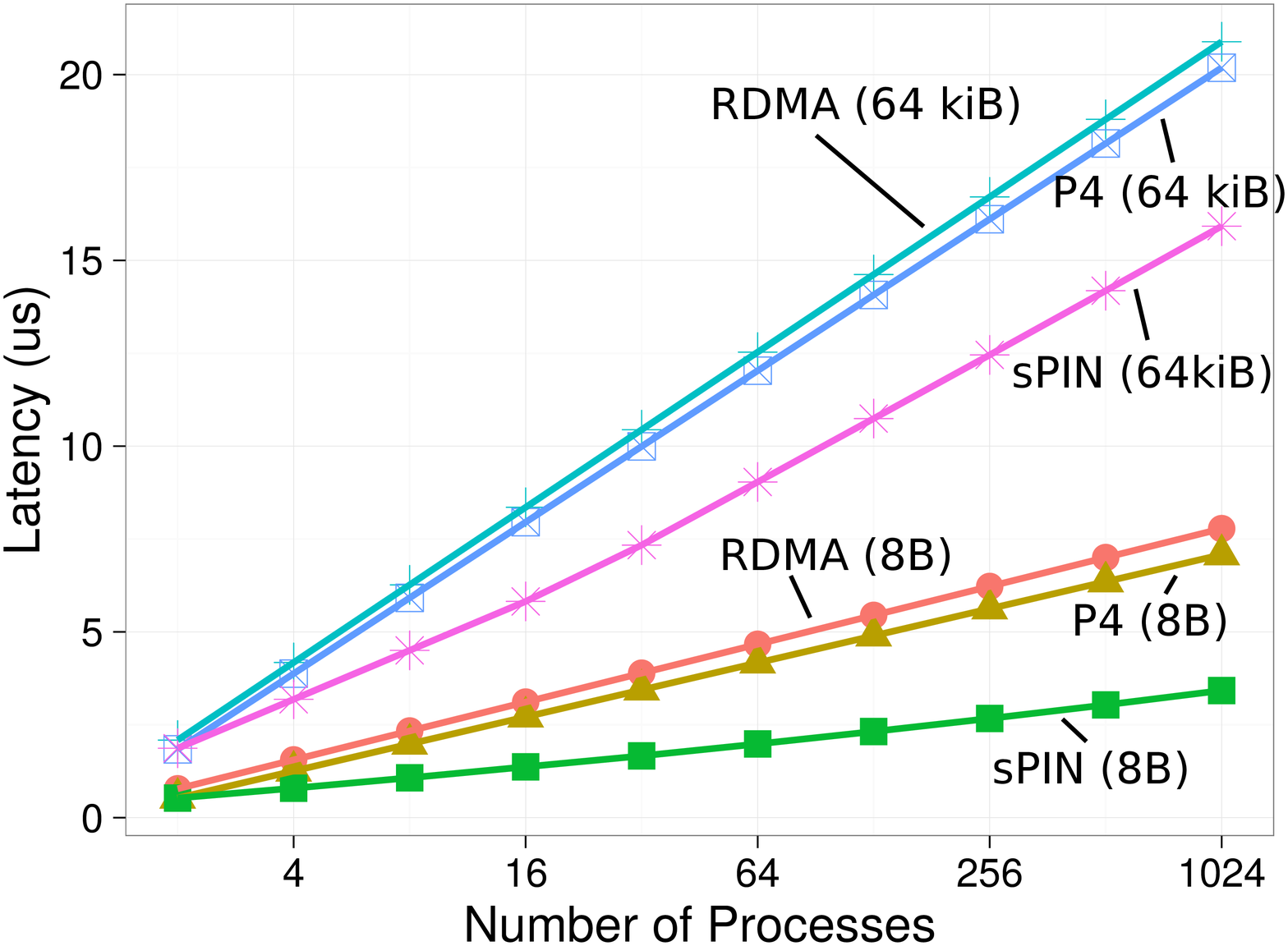} \label{fig:broadcast}
}%
    \hfill
    \subfloat[Matching protocols (left: small messages, right:
      large messages; top: recv called before arrival, bottom: after
        arrival)]{{
	\includegraphics[height=3.82cm]{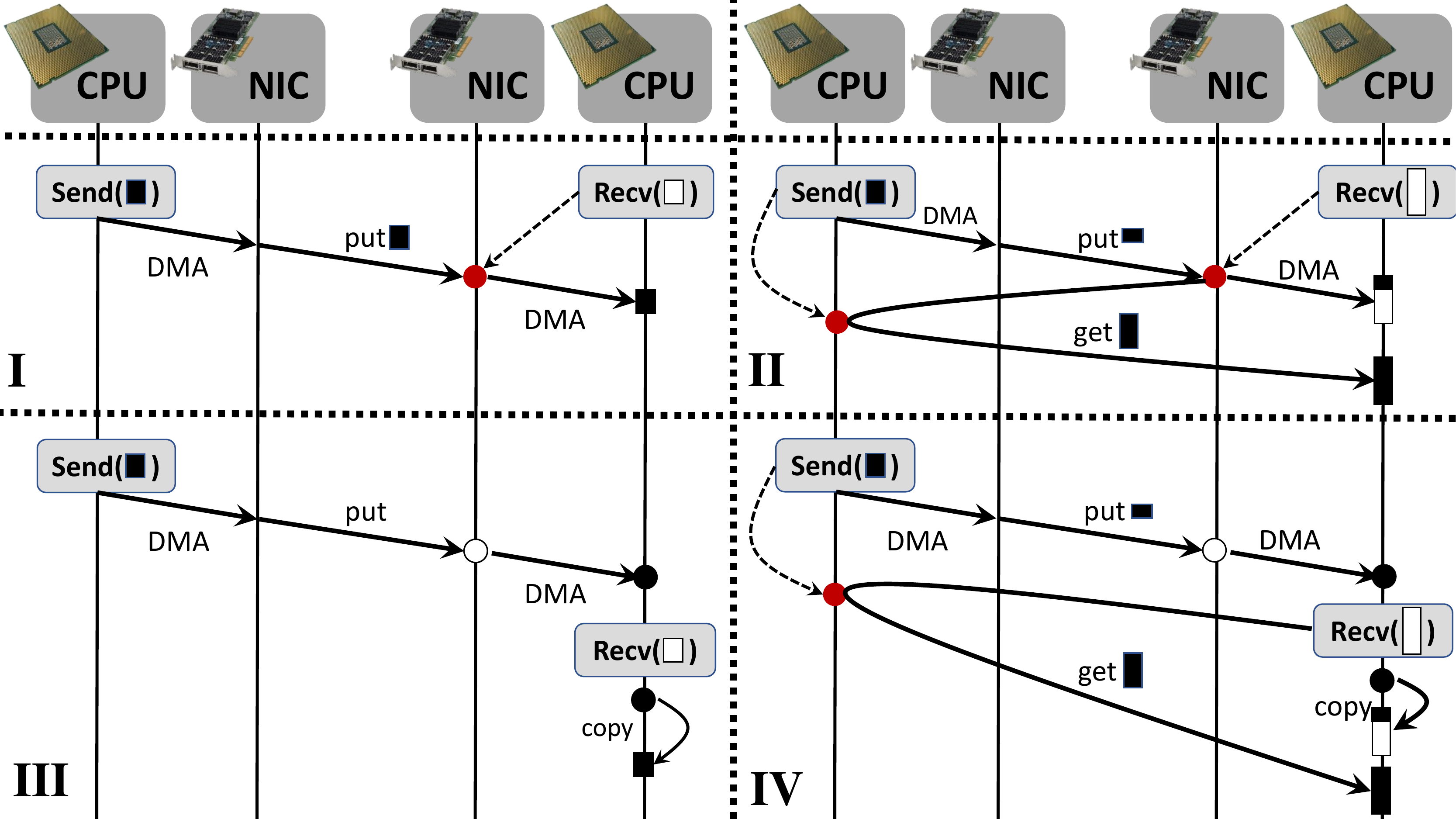}
  \label{fig:protocols}
}}%
    \hfill
    \subfloat[Application overview]{{
\footnotesize
\sf
  \begin{tabular}[b]{lllll}
\toprule
  \textbf{program} & \textbf{p} & \textbf{msgs}  & \textbf{ovhd} & \textbf{spdup}\\
\midrule
  MILC & 64 & 5.7M& 5.5\% & 3.6\%\\
  POP & 64 & 772M & 3.1\% & 0.7\%\\
  coMD & 72 & 5.3M & 6.1\% & 3.7\%\\
  coMD & 360 & 28.1M & 6.5\% & 3.8\%\\
  Cloverleaf & 72 & 2.7M & 5.2\% & 2.8\%\\
  Cloverleaf & 360 & 15.3M & 5.6\% & 2.4\%\\
\bottomrule
\end{tabular}
  \vspace{5em}
  \label{tab:apps}
}}%
  \vspace{-1em}
    \caption{Broadcast and message matching protocols implemented using
    \abbr{}}%
\end{figure*}
For our parameters, this means
$12.5\mathrm{Mmps} \leq \overline{\Delta}\leq 150\mathrm{Mmps}$ (Mmps
= million messages per second).
Figure~\ref{fig:hpusneeded} shows how many HPUs are needed to guarantee
line rate for different packet sizes and processing times.
With our design of 8 HPUs, we can support any packet size if the handler
takes less than $\hat{T}_s=53ns$. From $g/G=335$B, the link bandwidth becomes the
bottleneck and we can support full line rate as long as the handler
executes in less than $\hat{T}_l(s)=8Gs$. For full 4KiB packets,
$\hat{T}_l(4,096)=650ns$.

\subsubsection{\abbr{} Offloaded Broadcast}

For the last microbenchmark, we demonstrate the design of distributed
algorithms in \abbr{} using a broadcast operation. We implement a
binomial tree algorithm, which would require logarithmic space on a
Portals 4 NIC and would thus be limited in scalability. In \abbr{}, the
algorithm is not limited in scalability while it will occupy one HPU for
its execution.

We implemented the broadcast operation in RDMA on the CPU, in Portals 4
as predefined triggered operations, and with \abbr{} using
store-and-forward as well as streaming HPU kernels. As for ping-pong,
the store-and-forward mode sends messages that are smaller than a packet
directly from the device and from host memory otherwise. Thus, the
performance is always within 5\% of the streaming mode for single-packet
messages and to Portals 4 for multi-packet messages. Thus, we omit the
store-and-forward mode from the plots.

Figure~\ref{fig:broadcast} shows the small message (8 B) and
large-message (64 KiB) case 
for varying numbers of processes and the different implementations. We
observe the benefit of direct forwarding for small messages as well as
streaming forwarding for large messages. We only show data for the
discrete NIC configuration to maintain readability. The integrated NIC
has slightly lower differences but \abbr{} is still 7\% and 5\% faster
than RDMA and Portals 4 at 1,024 processes, respectively. The full handler
code is shown in Appendix~\ref{sec:broadcastcode}.

All benefits known from collective offloading
implementations~\cite{hemmert2010using,Roweth:2005:OGR:1104994.1105020,6008920}
such as asynchronous progression and noise-resilience remain true for
\abbr{}. As opposed to existing offloading frameworks that restrict the
collective algorithms (e.g., to pre-defined trees), \abbr{} supports
arbitrary algorithms (including pipeline and
double-tree~\cite{hoefler-moor-collectives}) due to the flexible programmability and high
forwarding performance of the HPUs. In fact, the very low overheads for
HPU packet processing suggest new streaming algorithms for collective
operations. We leave a more detailed investigation for future work.

\section{Use cases for \abbr{}}

We now discuss several more complex use cases for \abbr{}. The idea
is not to present full applications for which our cycle-accurate
simulation environment would be too slow, but to present detailed
simulation results of the critical pieces of real applications.

\subsection{Asynchronous Message Matching}

High-speed interconnection network attempts to offload as much work as
possible to the NIC. This is simple for remote memory access
(RMA) programming models where the source determines the destination
address and the NIC simply deposits the data into the correct virtual
address (as specified in the packet). 
However, this requires some form of distributed agreement on the
destination buffer offset. Message passing simplifies the program logic
and allows the target process to determine the local buffer location by
calling \lstinline$recv$. This simplification at the programming
level complicates the offloading of the matching because the correct
destination address may only be known after the packets of the message
arrive. Protocols to implement message passing over RDMA networks are
typically implemented by the
CPU~\cite{Woodall:2006:HPR:2091359.2091382}. However, progressing these
protocols requires synchronous interaction with the CPU. Thus,
communication/computation overlap for rendezvous as well as nonblocking
collective operations are often hindered~\cite{hoefler-ib-threads}. These issues led to the development of
specialized protocol offload engines that \abbr{} generalizes to a
universal network instruction set architecture (NISA).

Figure~\ref{fig:protocols} illustrates the matching process for small
messages (left) and large messages (right) as well as the cases
where the receive is posted before the message arrived (top) or after
the message arrived (bottom). 
The matching mechanism of Portals 4 and \abbr{} allows for offloading
progression and matching of small message transmissions. If the receive
is posted before the first packet arrives it installs a matching entry
(filled circle) and the NIC will deposit the data into the correct
memory at the target process upon receiving the message.  
Otherwise, as shown in case III, the packets will match a default action
(hollow circle) that stores the message into a predetermined location.
When the matching receive is then called later, the CPU finds the
message and copies the data into the receive buffer and completes
immediately.
This allows a data copy to be saved in case I, while RDMA will
always perform a copy (similar to case III).

If the data is too large to be buffered, the process is more complex
because it requires synchronization between the sender and the receiver.
Ideally, this is fully offloaded to the receiver's NIC (without
the need to synchronize on the receive-side CPU). 
Barrett et al.~\cite{barrett2011using} propose a protocol for Portals 4
where the receiver monitors the received bytes, and if a message arrives
that writes more than the eager threshold. This
message triggers a get to the source that is matched to the correct
pre-set-up memory. 
Unfortunately, this protocol is not practical due to the following
limitations: (1) it requires triggered gets to be set up for each of
the $P-1$ potential sources, requiring $\Omega(P)$ memory per process;
(2) it requires additional match-bits to keep a message counter
that is used to identify the correct matching entry at the source; and 
(3) it does not support wildcard receives (e.g.,
\lstinline$MPI_ANY_SOURCE$).

In \abbr{}, we implement a practical protocol that avoids all three
limitations as illustrated in the right half of Figure~\ref{fig:protocols}.
If the receive was called before the message arrived (case II), it sets
up a header handler and a payload handler for the first message (filled
circle at receiver).
The header handler checks whether the message is large or small
(determined by its size) and falls back to the normal Portals 4 handling
for small messages. If the message is large, the handler interprets the first and
second user-header as the total message size and the tag at the source.
Then, the header handler uses these two fields to issue a get operation
to the source. This get matches a descriptor that has
been set up during the send (filled circle at source). The payload
handler then deposits the payload of the message at the beginning of the
host's memory descriptor.  If the message arrived before receive was
called (case III) then the handler logic is executed by the main CPU.

The main benefits of \abbr{} compared to RDMA are one less copy in the
small-message case and completely asynchronous progress in the
large-message case. We simulate the influence of fully-offloaded
execution using LogGOPSim. The overhead of the local copy can be
significant because the network deposits data at a rate of 50 GiB/s,
while the local memory only delivers 150 GiB/s. This can lead to a copy
overhead of up to 30\%. We only considered point-to-point operations
for implementation because collective communication may use
specialized protocols. We simulated traces for several real-world and proxy
applications. Due to simulation overheads, we
only execute relatively short runs between 20 and 600 seconds. Yet, we
measure full application execution time including initialization and
output (from \verb@MPI_Init@ to \verb@MPI_Finalize@). Thus, we expect that the
speedups for longer runs are higher.  We discuss the results below and
summarize them in Table~\ref{tab:apps}.

\paragraph{MILC} The MIMD Lattice Computation (su3\_rmd) is used to study
Quantum Chromodynamics (QCD), the theory of the strong
interaction~\cite{bernard1991studying} as a finite regular hypercubic
grid of points in four dimensions with mostly neighbor interactions. 
We traced MILC on 64 processes where it spent 5.5\% of execution time in
point-to-point communications. In the simulation, MILC exchanged
5,743,212 messages and generated a total of 48M events. Fully offloaded
matching protocols improved the overall execution time by 3.6\%.

\paragraph{POP} The Parallel Ocean Program~\cite{jones2005practical}
(POP) models general ocean circulation and is used in several climate
modeling applications. The logically (mostly) rectangular problem
domain is decomposed into two-dimensional blocks with nearest-neighbor
communications and global exchanges. 
We traced POP on 64 processes where it spent 3.1\% of execution time in
point-to-point communications. In the simulation, POP exchanged
772,063,149 messages and generated a total of 1.5B events. Fully offloaded
matching protocols improved the overall execution time by 0.7\%.

\paragraph{coMD} The codesign app for molecular dynamics is part of the
Mantevo proxy application suite~\cite{heroux2009improving}. It features
the Lennard-Jones potential and the Embedded Atom Method potential.
We traced coMD on 72 processes where it spent 6.1\% of execution time in
point-to-point communications. In the simulation, coMD exchanged
5,337,575 messages and generated a total of 22M events. Fully offloaded
matching protocols improved the execution time by 3.7\%.

\paragraph{Cloverleaf} Cloverleaf is also part of the Mantevo proxy
applications~\cite{heroux2009improving} and implements a two-dimensional
Eulerian formulation to investigate materials under stress. 
We traced Cloverleaf on 72 processes where it spent 5.2\% of execution time in
point-to-point communications. In the simulation, Cloverleaf exchanged
2,677,705 messages and generated a total of 12M events. Fully offloaded
matching protocols improved the overall execution time by 2.8\%.

We remark that offloading message matching and asynchronous transmissions is not
limited to MPI. For example, Kim et
al.~\cite{Kim:2003:ETC:781498.781506} propose an asynchronous task
offloading model that could also be implemented with \abbr{}.

\subsection{MPI Datatypes}

Communicated data is often not consecutive in host memory. For example,
in a simple three-dimensional stencil code, only two of the six
communicated halos are consecutive in host memory. Most applications 
use process-local copying of the data to marshal it into a consecutive
buffer for sending and from a consecutive buffer for receiving. As
Schneider et al.~\cite{mpi-ddt-benchmark} point out, this is often not
considered as part of the communication overhead even though it is
in practicality, a part of the communication. Furthermore, they showed that
the data marshaling time can be up to 80\% of the communication time for
real-world applications because it is performed at both the send and
receive side. 

Data marshaling can be implemented in \abbr{} without the extra memory
copy, potentially reducing the communication overhead by 80\%. A
datatype processing library could implement this transparently to the
MPI user and upload a handler for each message. The HPUs would compute
the correct offsets on the NIC and DMA the data into the final location.
Here, \abbr{} not only improves the performance of the communication but
also relieves the memory subsystem of the host.

Without loss of generality, we focus on the most common strided access
that can be represented with MPI vector datatypes. Most of today's
networking interfaces (e.g., Portals 4 or OFED) support iovecs to
specify nonconsecutive accesses. However, they require $\mathcal{O}(n)$
storage to specify $n$ blocks to be copied, even for strided access.
With vector datatypes, strided access can be expressed as an
$\mathcal{O}(1)$ tuple of $\langle$\textit{start, stride, blocksize,
count}$\rangle$, where
count blocks of size blocksize are copied beginning from address start.

\begin{figure}[h!]
  \includegraphics[width=\columnwidth]{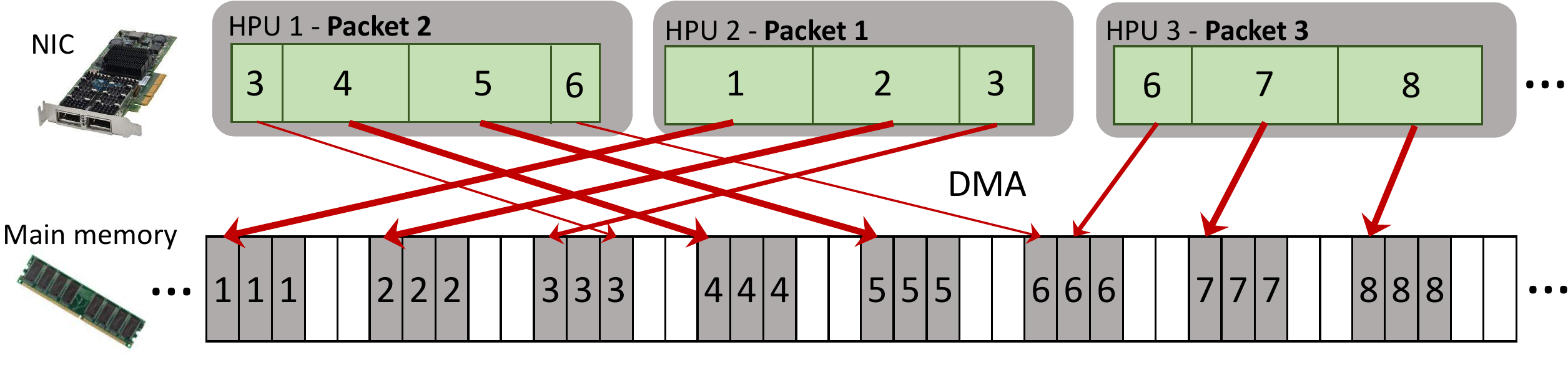}
  \vspace{-2.2em}
  \caption{Processing vector datatypes in payload handlers}
  \vspace{-1.2em}
  \label{fig:datatypes}
  \centering
\end{figure}
\begin{figure*}[t]%
    \centering
    \subfloat[Strided receive with varying blocksizes]{
  \includegraphics[height=3.62cm]{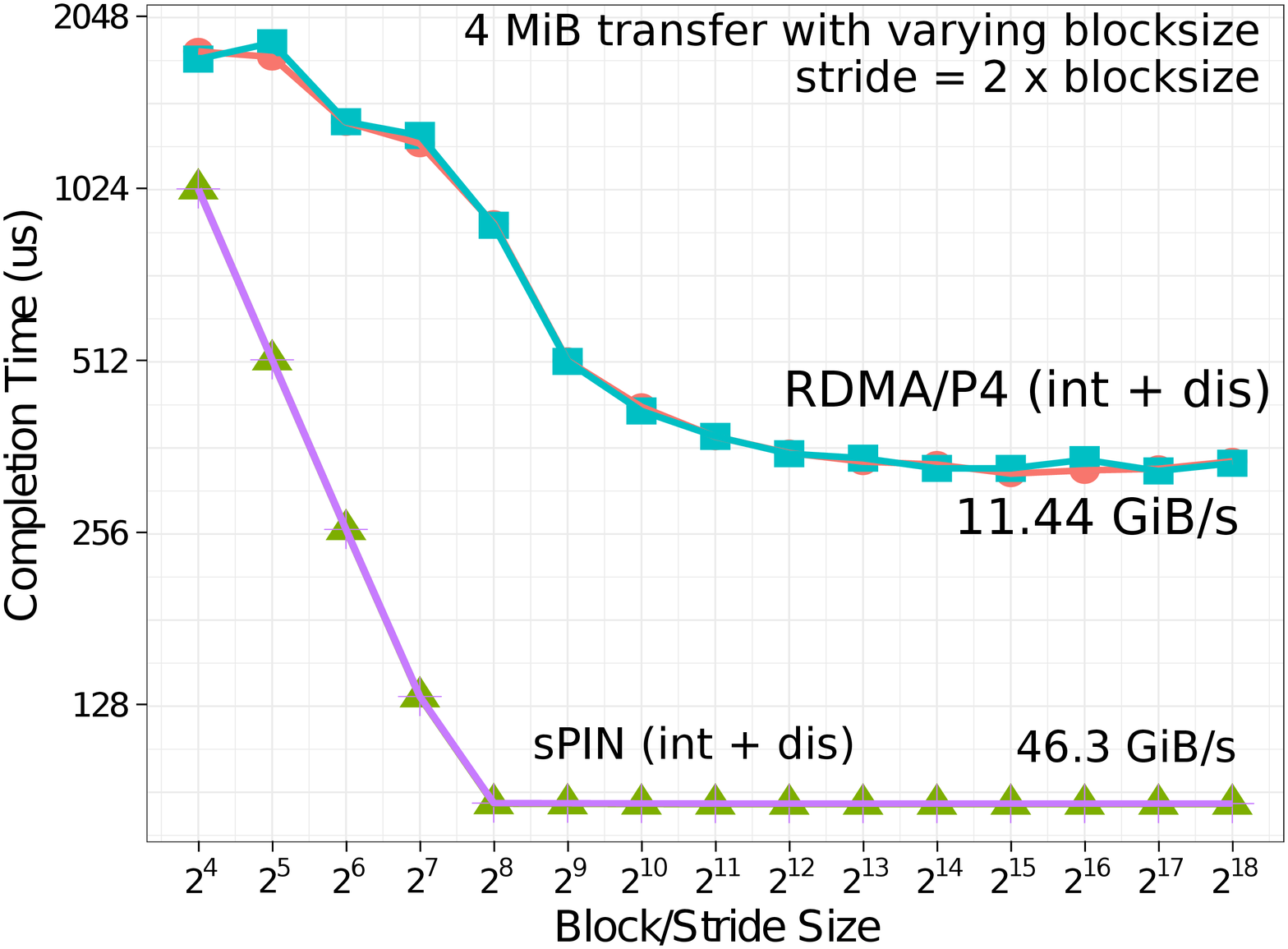}
  \label{fig:datatype_results}
}%
    \hfill
    \subfloat[Distributed RAID using RDMA (left) and \abbr{} (right)]{
	\includegraphics[height=3.62cm]{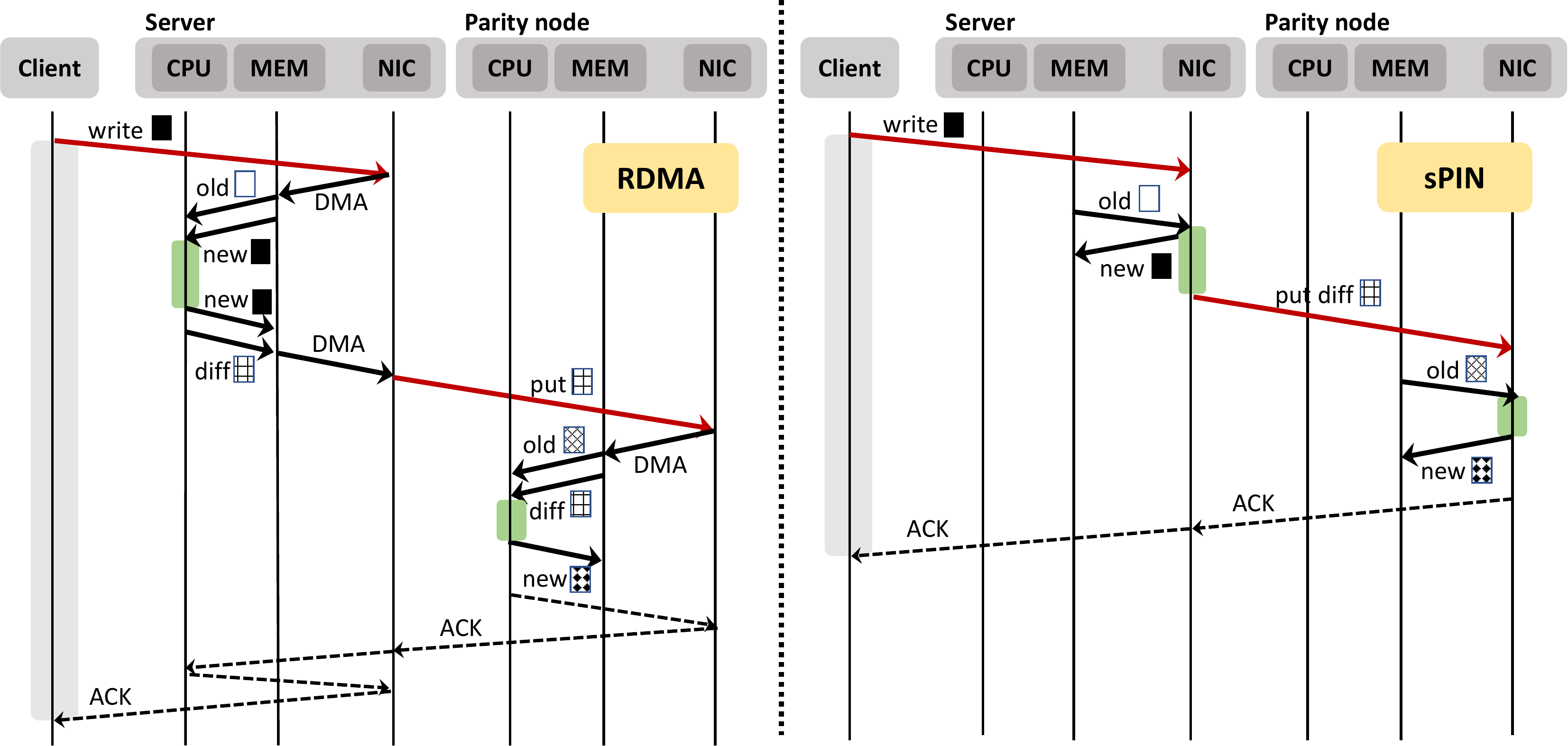}
  \label{fig:storage}
}%
    \hfill
    \subfloat[Update time in a distributed RAID-5 system]{
  \includegraphics[height=3.62cm]{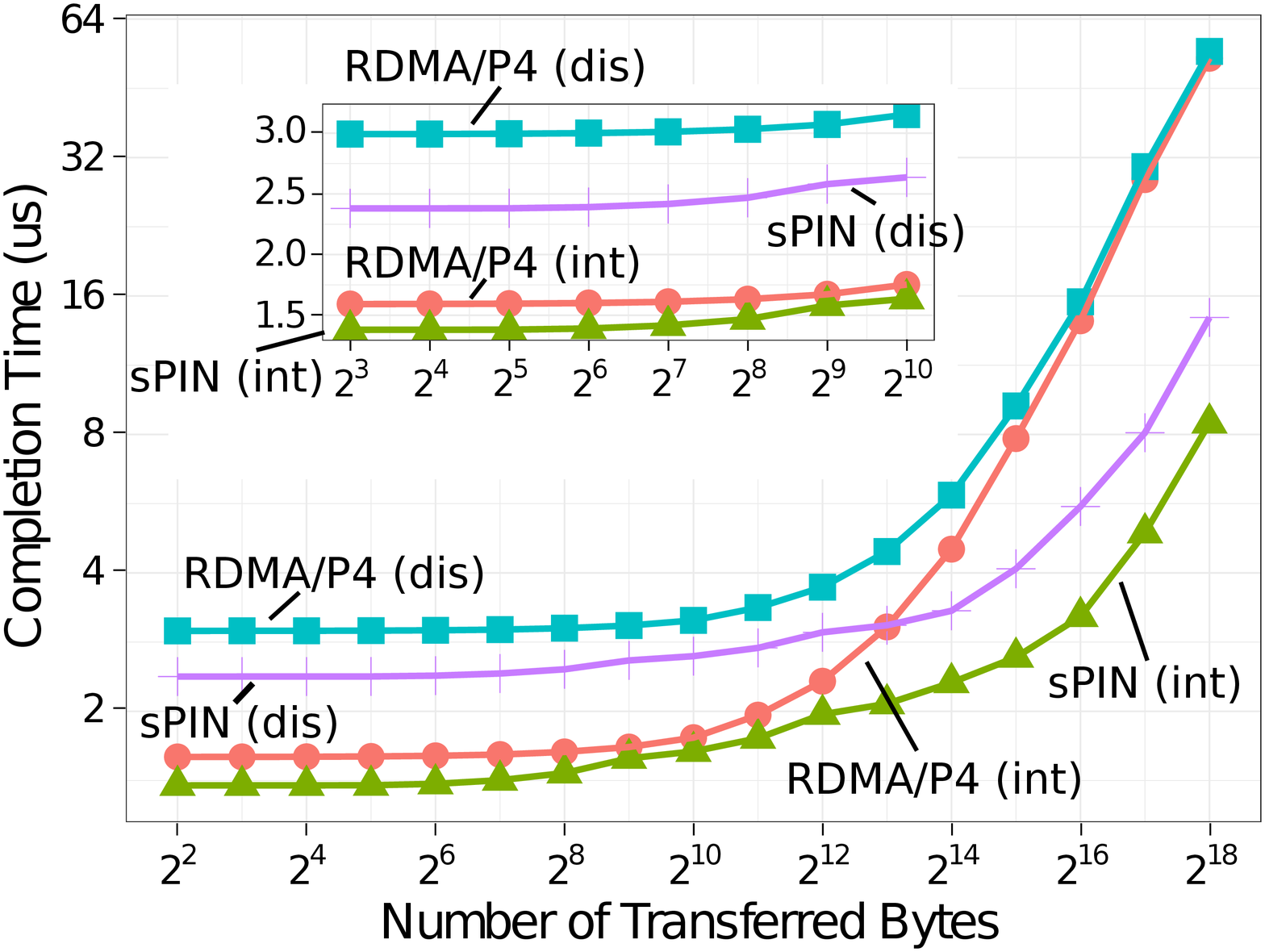} \label{fig:storage_data}
}%
  \vspace{-1em}
    \caption{Strided datatype and distributed RAID performance comparing
    RDMA/Portals 4 and \abbr{}}%
  \vspace{-1em}
\end{figure*}
Figure~\ref{fig:datatypes} illustrates how the payload handlers copy the
data into the correct positions for a strided layout. The figure shows
three packets of a stream that are deposited as a strided type into main
memory. The user only specifies the tuple $\langle$\textit{start, stride=2.5
KiB, blocksize=1.5 KiB, count=8}$\rangle$ and the MTU is 4 KiB (the
illustration omits headers for clarity). The packets can be processed by
the payload handlers in any order or in parallel because each packet
carries its offset in the message and the payload handlers computes the
correct offset in the block. Arrows represent DMA writes and their width
indicates the size of the transaction. The full code is shown in
Appendix~\ref{sec:datatypecode}. 

To demonstrate the performance of \abbr{} in practice, we simulate an
execution of datatype processing (unpack) at the destination. For this,
we choose a fixed message size of 4 MiB and vary the blocksize. We
keep the strides fixed at twice the blocksize.
Figure~\ref{fig:datatype_results} shows the results. The DMA overhead
for small transfers dominates up to block size 256, then \abbr{} is able
to deposit the data nearly at line-rate (50 GiB/s) while RDMA remains at a
bandwidth around 8.7 GiB/s due to the additional strided copies.

\subsection{Distributed RAID Storage}

After demonstrating \abbr{}'s benefits for parallel applications, we now
show that it can also benefit system software on large-scale compute
systems. For example, filesystems often use replication at the object
storage layer involving multiple nodes to improve reliability of the
data~\cite{weber2007high}. 
We assume that the inode-to-block lookup has been performed by the
client that addresses the correct blocks at the storage server. 
Blocks are accessed through the network and if a block is updated, the
parity block on another server needs to be updated as well. The
computation is simple: $p'= p \oplus n' \oplus n$ where $p$ and $p'$ are the
old and new parity blocks and $n$ and $n'$ are the old and new data
blocks, respectively.

Filesystem nodes are often accessed through RDMA (e.g.,
object data servers in Lustre). Since replication is totally transparent
to the client, RDMA cannot be used directly because it would reply
before the parity node is updated. Thus, such systems implement a more
complex protocol using the target's CPU as shown in the left part of
Figure~\ref{fig:storage}. With \abbr{}, the server NIC can issue
requests to update the parity without involving the servers CPU.
Furthermore, the parity node's NIC can apply the update in host memory.
This can easily be used to implement a traditional disk-backed storage
server or a distributed memory system similar to
ramcloud~\cite{ousterhout2010case}. In both cases, \abbr{} offloads most
of the storage protocol to the NIC.

We show a simple latency/bandwidth benchmark comparing in-memory
storage using four data nodes and one parity node in a RAID-5 configuration.
For this test, we update contiguous memory of growing size strided
across the four data nodes and measure the time until all
ACKs are received (after updating the parity node).  Figure~\ref{fig:storage_data}
shows the performance comparing RDMA and \abbr{}. The results
demonstrate the comparable performance for small messages and the
significantly higher bandwidth of \abbr{} for large block transfers, the
common case for parallel filesystems. 
To demonstrate \abbr{}'s influence on real-world workloads, we simulate
five traces obtained from the Storage Performance
Council~\cite{spc-traces}. The first two traces represent OLTP applications
running at a large financial institution. The remaining three are I/O
traces from a popular search engine. \abbr{} improves the processing
time of all traces between 2.8\% and 43.7\%. The integrated \abbr{} NIC
with financial traces showed the largest speedup. 
The full handler code is shown in Appendix~\ref{sec:reedsolomoncode}.

\subsection{Other Use Cases}

We have demonstrated all key features of \abbr{} in the previous
sections. However, many applications, tools, and system services can
benefit from \abbr{}. Here, we outline other use cases for which we have
to omit a detailed analysis due to space restrictions.

\paragraph{Distributed Key-Value Stores}

Distributed key-value stores provide a storage infrastructure where data is
identified by keys that simplify the application's storage
interface~\cite{fitzpatrick2004distributed}. 
They can be compared to large distributed two-level hash-tables. The
first level determines the target node and the second level determines a
location at that node. Let $(k,v)$ represent a key-value pair with $k\in
\mathcal{K}$ and $v\in \mathcal{V}$. We assume that there exist two hash
functions $H_1(x): \mathcal{K} \mapsto \{0..P-1\}$ and $H_2(x):
\mathcal{K} \mapsto \{0..N-1\}$ where $P$ and $N$ are the number of
nodes and hashtable-size per node, respectively.  We assume $H_1$ and
$H_2$ are reasonably balanced with respect to the expected key
distribution but not generally free of conflicts. 
Various high-performance RDMA implementations with varying complexity exist,
ranging from replicated state
machines~\cite{Poke:2015:DHS:2749246.2749267} and HPC storage
layers~\cite{Docan:2010:DIC:1851476.1851481} to distributed
databases~\cite{Dragojevic:2014:FFR:2616448.2616486,Dragojevic:2015:NCD:2815400.2815425}.

We now describe how \abbr{} could be used to offload the insert
function: A client that wants to insert the KV pair $(k,v)$ first
computes $H_1(k)$ to determine the target node $p$. Then, it computes
$H_2(k)$ and crafts a message $(H_2(k),len(k),k,v)$ (where $len(k)$ is
the size of the key in Bytes) to be sent to node $p$. 
We use a header handler to allocate memory to deposit $v$ and link it to
the correct position $H_2(k)$ in the hash table. Depending on the hash table
structure (e.g., closed or open), the handler may need to walk through a
list in host memory. To not back up the network, the header handler
would abort after a fixed number of steps and deposit the work item to
the main CPU for later processing. Other functions such as get or delete
can be implemented in a similar way. 

\paragraph{Conditional Read}

Many distributed database problems scan remote tables using simple
attributes. For example the statement \texttt{SELECT name FROM employees WHERE
id = 100} may cause a full scan of the (distributed) table
\texttt{employees}.  Reading all data of this table via RDMA would be a
waste of network bandwidth.  
Since our current handler definition does not allow interception and
filtering of the data for a get operation, we implement our own request-reply protocol.
The request message contains the filter criterion and a memory range and
the reply message contains only the data that matches. The more complex
query offload model described by Kumar et
al.~\cite{Kumar:2006:EPN:1898953.1899010} can also be
implemented in \abbr{}.

\paragraph{Distributed Transactions}
Distributed transactions require bookkeeping of the addresses that have
been accessed during the transaction. Complex protocols using memory
protection mechanisms are employed for multi-core
CPUs~\cite{Herlihy:1993:TMA:173682.165164}. We can use \abbr{} to log
remote accesses to local memory. For this, we introspect the header
handlers of all incoming RDMA packets and record the accesses in main
memory. The introspection can be performed at line rate and transaction
management is then performed at commit time on the host by evaluating
the logs. 

\paragraph{Simple Graph Kernels}
Many graph algorithms have very simple functions to be invoked for each
vertex. For example, a BFS only checks if the vertex was not visited
before and assigns it a number at the first visit. Shortest
path search (SSSP) algorithms update a vertex' distance with the minimum
of its current distance and the preceding vertex' distance plus
the weight of the connecting edge.

In distributed settings, node-boundaries can be crossed by
the traversal. Then, messages are sent from the originating vertex
to the destination vertex on the remote node. A message contains the
new distance (i.e., the distance of the source vertex plus the edge
weight). The remote handler then (atomically) checks if the
destination vertex needs to be updated and conditionally performs the update.

This is typically implemented by receiving messages in batches into
buffers and processing them on the main CPU. Yet, this requires to store
and load the message data from and to memory just to discard it after
the update. With \abbr{} we can define an offloaded handler to process
the updates immediately and save memory bandwidth.

\paragraph{Fault-tolerant Broadcast}
There are many different ways to implement a fault-tolerant broadcast.
Some rely on failure detectors and a two-phase broadcast-and-check
protocol, where the root restarts the broadcast with a different tree
if nodes have failed~\cite{buntinas2012scalable}. Others 
redundantly send messages in a virtual topology such as a binomial
graph~\cite{angskun2007binomial}. The former rely on failure detectors,
which cannot easily be implemented in the current RDMA or Portals 4
networks. The latter guarantee reliable delivery for less than
$\log_2 P$ failures and often outperform the broadcast-and-check protocols
in practice. 

Usually, these protocols are implemented with the help of the main CPU
to forward messages. This means that all $\log_2 P$ redundant messages
must be delivered to host memory. We can use \abbr{} to accelerate such
protocols by only delivering the first message to the user. This would
enable a transparent reliable broadcast service offered by the network.

\section{Related work}

We already discussed the relation of \abbr{} and active messages and
switch-based packet processing such as P4 in
Section~\ref{sec:background}. Programmable NICs have existed in
special-purpose environments. For example, Quadrics
QSNet~\cite{petrini2002quadrics} offered a programming interface that
was used to offload collectives~\cite{1303191} and an early Portals
implementation~\cite{pedretti-offload} used a programmable NIC. 
However, QSNet had to be
programmed at a very low level and was rather limited, which hindered
wider adoption. Myrinet provided open firmware that allowed researchers to
implement their own modules on the specialized NIC cores in C~\cite{1392618}.
As opposed to these constrained solutions, \abbr{} defines a simple
offload interface for network operations, similar to the ones that have
widely been adopted for compute offloading.

High-speed packet processing frameworks used for router
implementations, software defined
networking~\cite{1392618}
and P4~\cite{bosshart2014p4} provide similar functions. They also relate
well to \abbr{} in that the key idea is to apply simple functions to
packets.  However, these frameworks are not designed to interact with
host memory and the execution units are stateless and are thus much less
powerful than \abbr{}.

\section{Discussion}

\paragraph{Will \abbr{} NICs be built?}

With \abbr{}, we define an offloading interface for NICs (which we call
NISA) and we outline the requirements for a NIC microarchitecture. Using
our results from simulating ARM CPUs, a vendor could immediately build
such a NIC. In fact, we are aware of several vendors that will release
smart NICS that can be programmed to support \abbr{} with a similar
microarchitecture this year. \abbr{} enables the development of a
vendor-independent ecosystem just like MPI where third parties can
develop libraries for domain-specific handlers. Network acceleration
could then be, very much like NVIDIA's cuBLAS or NCCL libraries, 
offered for domains outside of HPC, such as machine learning and
data analytics, to impact bigger markets. 

\paragraph{Can \abbr{} work with other libraries than Portals 4}
Yes, for example, it would be straight-forward to define \abbr{}'s
handlers for OFED or Cray's uGNI. Here, the three handlers would not be
attached to a matching entry but a queue pair and they would be invoked
for every arriving message. One can also define \abbr{} for
connection-less protocols such as SHMEM or Cray's DMAPP. Here, one would
define handlers to be executed for messages arriving at certain contexts
or address ranges (similar to an ME). We chose to demonstrate the
functionality with the most complex interface, Portals 4, the principles
remain the same for others.

\paragraph{Can \abbr{} be executed on network switches?}
The definition of handlers allows them to be executed at any switch in
the network with some limitations. Since they're not associated with a
host, the functions put and get from host are not allowed and DMA
commands cannot be issued. Yet, the handlers can manipulate packets and
use their shared memory to keep state. We believe that this extension
would be simple but we defer a detailed analysis of use cases to future
work.

\paragraph{What if \abbr{} handlers run too long?}
In general, handlers may run for a very long time and incorrect handlers
may even not terminate. We would recommend to kill handlers after a
fixed number of cycles and move the interface into flow control.
However, this flow-control behavior is specific to Portals 4. In
general, one can imagine various ways to deal with slow handlers. We do
not recommend backing-up data into the (most likely lossless)
interconnect because a bad handler may block the whole network. Instead, 
arriving packets that cannot be processed can be dropped and
the user, once notified of this event, can tune the handlers until they can
perform at line-rate.

\section{Summary and Conclusions}
\balance

We defined \abbr{}, a vendor-independent and portable interface for
network acceleration. We discuss a reference implementation for Portals
4 and develop and share a simulation infrastructure that combines a
network and a microarchitecture simulator to analyze the benefits of
network offloading\footnote{https://spcl.inf.ethz.ch/Research/Parallel\_Programming/sPIN}. 
We show several use cases of how it can be used in
real-world parallel applications as well as system services for data
management. Our simulations demonstrate significant speedups for real
applications as well as important kernels.

We believe that \abbr{} will change the way we approach networking and
how we design NICs in the future---it will make exposing the specialized
data movement cores on the NIC to the user simple and enables the
development of a sophisticated ecosystem. 

\subsection*{Acknowledgments} TH edited the manuscript and developed the
original idea and specified the interface with input from RB and RG. SG
developed the simulation toolchain and implemented the first prototype.
KT implemented the handler codes and performed all experiments. 

We thank Keith Underwood, James Dinan, Charles Giefer, and Sayantan Sur
for helpful discussions during the initial design. The interface and
theory was developed during a research stay of the first author at
Sandia National Laboratories.

\bibliographystyle{ACM-Reference-Format}
\bibliography{sigproc}

\clearpage
\newpage
\appendix

\section{Artifact Description Appendix: sPIN: High-performance streaming
Processing in the Network}

\subsection{Abstract}

The results presented in this paper are generated with a simulation
tool-chain based on the cycle-accurate gem5 simulator and the
packet-level LogGOPSim simulator. The whole tool-chain is 
available to the public. 

\subsection{Description}

\subsubsection{Check-list (artifact meta information)}

\begin{description}
  \item [Program] LogGOPSim and gem5
  \item [Compilation] Standard GNU Linux environment (tested on Debian,
    Ubuntu and Redhat)
  \item [Binary]  Available as source-code only (for transparency).
    Build instructions are included
  \item [Data-set] Small ones ($<$100MiB) are included, larger ones
    ($>$100 GiB) upon request, see below.
  \item [Hardware] Execution tested on x86, simulated hardware ARMv7.
  \item [Publicly available]: Yes, see below.
\end{description}

\subsubsection{How software can be obtained (if available)}

We release the whole simulation tool-chain with a mini-howto and
instructions to reproduce the data in the paper on our webpage: https://spcl.inf.ethz.ch/Research/Parallel\_Programming/sPIN. 
After downloading and unpacking
the tar-ball, follow the instructions in the README
to build the two simulators. The README also describes part of the
software infrastructure. The directory also contains a README\_SC17,
which contains instructions how to generate the data used in this paper.

\subsubsection{Hardware dependencies}

None

\subsubsection{Software dependencies}

Standard gem5 installation, agraph and cgraph as included in default
Debian/Ubuntu/Redhat systems. See README in the package for details.

\subsubsection{Datasets}

All but one are included, the large ($>$100 GiB is available on demand),
see above.

\subsection{Installation}

See README in package.

\subsection{Experiment workflow}

See README\_SC17 in package.

\subsection{Evaluation and expected result}

The exact results we show in the paper can be reproduced using the
scripts and simulators we used. 

\subsection{Experiment customization}

None

\subsection{Notes}

In the following, we provide a detailed specification of the P4\abbr{}
interface so that readers can follow the details and implement a \abbr{}
system. Furthermore, in Appendix~\ref{app:code}, we provide the source
code of all handlers described in this paper for convenience. All source
codes are also included in the tar-package linked above.

\section{Detailed P4sPIN Interface}
Here, we describe the detailed C interface for P4sPIN to ensure
reproducibility.

\subsection{Detailed MD descriptor}\label{sec:p4md}
\begin{lstlisting}
ptl_me_t { 
   ...  // original Portals 4 arguments

   ptl_handler_t header_handler;
   ptl_handler_t payload_handler;
   ptl_handler_t completion_handler;

   ptl_hpu_md_h  hpu_memory;
   void *hpu_initial_state;
   ptl_size_t hpu_initial_state_length;

   void *handler_host_mem_start;
   ptl_size_t handler_host_mem_length;
}
\end{lstlisting}
\sloppy
The three handlers can be installed independently. The user can choose
to not have a specific handler called by setting it \lstinline$NULL$.
To keep the size of the \lstinline$ptl_me_t$ struct small if no handlers
are needed, one could specify a sub-struct \lstinline$ptl_handler_data$
for all these additional elements. This sub-struct could be set to
\lstinline$NULL$ if no handlers are installed.  

\subsection{HPU Memory Management}\label{sec:hpumem}

\begin{lstlisting}
int PtlHPUAllocMem(ptl_size_t length, ptl_handle_ni_t ni, ptl_hpu_md_h *hpu_mem)
int PtlHPUFreeMem(ptl_hpu_md_h *hpu_mem)
\end{lstlisting}
This call allocates length memory in device \lstinline$ni$ and stores all
information in the opaque handle \lstinline$hpu_mem$, which is then
associated with an \lstinline$ptl_me_t$.

\hpara{handler initial state}
Handlers can read and write host memory and parameters could be passed
through this memory. Yet, it is often useful to initialize HPU memory
with some small control values that are set from the host while
installing the ME and handler. When
posting an ME, the host can specify a memory region with
\lstinline$hpu_initial_state$ that is used to
initialize that HPU memory. This feature can be used to coordinate multiple header
handlers working on the same HPU memory.
The length is the size of the pre-allocated
state memory that is passed to the header handler; it must be smaller
than \lstinline$max_initial_state$ (cf.~Section~\ref{sec:limits}). The state will always be allocated
but only be overwritten if \lstinline$hpu_initial_state$ is not
\lstinline$NULL$. 

\hpara{host spaces}
\sloppy
The ME identifies host memory to steer the access to. If a handler is
present, it may be useful to have additional memory for the handler 
data (e.g., to collect statistics about messages).  The fields
\lstinline$handler_host_mem_start$ and
\lstinline$handler_host_mem_length$ identify a second range of host memory
where the handler can store its output. An additional option
\lstinline$HANDLER_IOVEC$ specifies if the handler memory start and
length are to be interpreted as an iovec. 

\hpara{pending MEs}
A handler can generate messages from the context of an ME that change
the completion semantics of an ME. For example, an MPI rendezvous
message may arrive at a posted rendezvous handler, which then in turn
posts a get operation to fetch the data. In this case, the handler can
return \lstinline$PENDING$ which instructs the runtime to not complete the
ME once this message is processed but wait for another message that
matches it. 

\subsubsection{NI Limits}
\label{sec:limits}
All resources in Portals 4 are strictly limited to allow for an
efficient hardware implementation. The available resources for each logical
network interface (NI) are reflected in the limits structure. To support
\abbr{}, we add the following fields:

\vspace{1em}
\setlength{\tabcolsep}{3pt}
\hspace*{-1.3em}
\begin{tabular}{l p{4.6cm}}
\textbf{parameter name} & \textbf{description} \\
\hline
\lstinline$max_user_hdr_size$       & maximum size of a user-header at a packet
                            (maximum size the user can add to the header, 
                            for quick parsing) \\
\hline
\lstinline$max_payload_size$        & maximum size of payload data in a packet (used
                            to determine requirements for payload handler) \\
\hline
\lstinline$max_handler_mem$         & maximum bytes of HPU memory for a handler \\
\hline
\lstinline$max_initial_state$         & maximum bytes of HPU initial state for a handler \\
\hline
\lstinline$min_fragmentation_limit$ & minimum allowed unit (bytes) for packet processing
                            (each payload handler's data is always 
                            guaranteed to be naturally aligned to this 
                            limit as well as be a multiple of this
                            limit, a high-quality implementation makes
                            this as big as possible)\\
\hline
\lstinline$max_cycles_per_byte$    & maximum number of HPU cycles per byte payload \\
\hline
\end{tabular}

\subsection{Header Handler Details}

The pointer state points at the initial data in HPU memory. This data
may have been initialized by the host during installation of the ME. 
The struct \lstinline$ptl_header_t$ contains the following elements:
\begin{lstlisting}
struct ptl_header_t {
  ptl_request_type_t type; // put, get, atomic
  ptl_size_t length; // payload length
  ptl_process_t target_id; // target nid/pid
  ptl_process_t source_id; // source nid/pid
  ptl_match_bits_t match_bits; // match tag
  ptl_size_t offset; // offset in ME
  ptl_hdr_data_t hdr_data; // inline data
  ptl_user_header_t user_hdr; // user header
}
\end{lstlisting}
The struct \lstinline$user_header_t$ is user-defined and can be
used by the compiler and HPU to parse headers. It allows access to the first
bytes of the payload of the header message as user-defined header
structures. 

The return code of the handler is used to influence the runtime system.
We define the following return codes:

\vspace{1em}
\setlength{\tabcolsep}{3pt}
\hspace*{-1.3em}
\begin{tabular}{ l  p{5.0cm}  }
\hline
\lstinline$DROP$ & handler executed successfully and message shall be
       dropped (NIC will discard all following packets)\\
\hline
\lstinline$DROP_PENDING$ & same as \lstinline$DROP$ but do not complete ME\\
\hline
\lstinline$PROCESS_DATA$ & handler executed successfully, NIC shall
               continue calling payload handlers for packets\\
\hline
\lstinline$PROCESS_DATA_PENDING$ & same as \lstinline$PROCESS_DATA$ but do not complete ME\\
\hline
\lstinline$PROCEED$ & handler executed successfully, NIC shall execute
          the default action identified by the request and not invoke
          any further handlers. If the default action is to deposit the
          payload at the ME offset, then this payload will include the
          user-header. \\
\hline
\lstinline$PROCEED_PENDING$ & same as \lstinline$PROCEED$ but do not complete ME\\
\hline
\lstinline$SEGV$ (*) & segmentation violation\\
\hline
\lstinline$FAIL$ (*) & handler error (user-returned)\\
\hline
\end{tabular}
\vspace{1em}

Return codes marked with (*) are considered errors and will raise an
event in the event queue associated with the ME. If multiple errors
occur while processing a message, only the first one is reported in the
event queue.

\subsection{Payload Handler Details}

The pointer state points at the initial data in HPU memory. This data
may have been initialized by the host or header handle. 
The struct \lstinline$ptl_payload_t$ contains information about the
payload data:
\begin{lstlisting}
struct ptl_payload_t {
  ptl_size_t length; // length of the data
  ptl_size_t offset; // payload offset in message
  uint8_t base[0]; // beginning of data
}
\end{lstlisting}

The return code of the handler is used to influence the runtime system.
We define the following return codes:

\vspace{1em}
\setlength{\tabcolsep}{3pt}
\hspace*{-1.3em}
\begin{tabular}{ l  p{6.0cm}  }
\hline
\lstinline$DROP$ & handler executed successfully, drop packet\\
\hline
\lstinline$SUCCESS$ & handler executed successfully \\
\hline
\lstinline$FAIL$ (*) & handler error (user-returned)\\
\hline
\lstinline$SEGV$ (*) & segmentation violation\\
\hline
\end{tabular}
\vspace{1em}

Return codes marked with (*) are considered errors and will raise an
event in the event queue associated with the ME. If multiple errors
occur while processing a message, only the first one is reported in the
event queue. Note that ``first'' may not be well defined for payload
handlers because they may execute in parallel.

\subsection{Completion Handler Details}

The return code of the handler is used to influence the runtime system.
We define the following return codes:

\vspace{1em}
\setlength{\tabcolsep}{3pt}
\hspace*{-1.3em}
\begin{tabular}{ l  p{5.9cm}  }
\hline
\lstinline$SUCCESS$ & handler executed successfully \\
\hline
\lstinline$SUCCESS_PENDING$ & same as \lstinline$SUCCESS$ but do not complete ME!\\
\hline
\lstinline$FAIL (*)$ & handler error (user-returned)\\
\hline
\lstinline$SEGV (*)$ & segmentation violation\\
\hline
\end{tabular}
\vspace{1em}

Return codes marked with (*) are considered errors and will raise an
event in the event queue associated with the ME. If multiple errors
occur while processing a message, only the first one is reported in the
event queue. If either the header or completion handler returned a
\lstinline$PENDING$ code then the ME will not be completed after
completing the message. 

  \subsection{Handler Actions Details}\label{sec:hdlract}
\hpara{DMA}
We specify blocking and nonblocking DMA calls to allow the HPU to copy
to/from host memory. The blocking DMA calls block the HPU thread until
the data arrived (the HPU can context-switch to another thread) and the
nonblocking DMA calls return a handle for later completion. Nonblocking
DMA calls are slightly higher overhead due to handle allocation and
completion and should only be used if the DMA can be overlapped with
other HPU instructions. Blocking DMA requests can be seen as an
indication of urgency and could be prioritized by the DMA subsystem. We
show the nonblocking interfaces below. 
\begin{lstlisting}
int PtlHandlerDMAToHostNB(const void *local, ptl_size_t offset, ptl_size_t len, unsigned int options, ptl_dma_h *h);
\end{lstlisting}
This function copies len bytes from local to offset in ME. Options can
either set \lstinline$PTL_ME_HOST_MEM$ or
\lstinline$PTL_HANDLER_HOST_MEM$ to select the host memory space). 
\begin{lstlisting}
int PtlHandlerDMAFromHostNB(ptl_size_t offset, void* local, ptl_size_t len, unsigned int options, ptl_dma_h *h);
\end{lstlisting}
\sloppy
This function copies len bytes from offset in ME to local memory.
Options can either set \lstinline$PTL_ME_HOST_MEM$ or
\lstinline$PTL_HANDLER_HOST_MEM$ to select the host memory space).
Both functions return immediately 
with a
handle that can be used to check for completion. The blocking interfaces
\lstinline$PtlHandlerDMAToHostB()$ and
\lstinline$PtlHandlerDMAFromHostB()$ accept the same arguments but do
not return a handle.

\hpara{DMA atomics}
Standard DMA calls to and from the host are not atomic. Atomic DMAs over
PCI can be expensive (and might require locking the ME explicitly),
thus, we offer an atomic DMA compare-and-swap function for
synchronization. Data has to be naturally aligned:
\begin{lstlisting}
int PtlHandlerDMACASNB(ptl_size_t offset, uint64_t *cmpval, uint64_t swapval, unsigned int options, ptl_dma_h *h);
\end{lstlisting}
This call compares the value at the offset location with cmpval and
replaces it with swapval if they are equal.  If the CAS fails, cmpval
is overwritten with the value at the offset location. 
\begin{lstlisting}
int PtlHandlerDMAFetchAddNB(ptl_size_t offset, ptl_size_t inc, ptl_size_t *res, ptl_type_t t, unsigned int options, ptl_dma_h *h);
\end{lstlisting}
This call atomically increments the value of type t at location offset
by inc and returns the value before in res.
Similar blocking function exist for both atomic calls.

\hpara{DMA completion}
The nonblocking completion test function 
\begin{lstlisting}
int PtlHandlerDMATest(ptl_dma_h handle);
\end{lstlisting}
returns true if handle is complete and the data transfer is finished,
false otherwise. 
The completion wait function 
\begin{lstlisting}
int PtlHandlerDMAWait(ptl_dma_h handle);
\end{lstlisting}
waits until handle is complete and the data transfer is finished. A
handle can be re-used only after it has been completed. 

\hpara{thread yield}
Multiple HPU threads may share the same HPU memory and run
simultaneously. We define a compare-and-swap function on HPU memory to
provide powerful synchronization features:
\begin{lstlisting}
int PtlHandlerCAS(uint64_t *value, uint64_t cmpval, uint64_t swapval);
\end{lstlisting}
The function atomically tests value and cmpval for equality and replaces
value with swapval if the test succeeds. It returns true if the swap
succeeded and false otherwise.
Furthermore, we define a fetch-and-add function for HPU threads:
\begin{lstlisting}
int PtlHandlerFAdd(uint64_t *val, uint64_t *before, uint64_t inc);
\end{lstlisting}
This function atomically increments the value val with inc and returns
the value before.
HPU threads are not de-scheduled automatically but they can yield
the HPU voluntarily using the function call
\lstinline$PtlHandlerYield()$. A runtime environment is free to ignore
this call but it can be understood as a hint to schedule another thread.
For example, when waiting for a DMA, yielding allows the HPU to use its
processing resources more efficiently than polling. 

\hpara{messaging}
Handlers can send messages either from HPU or host memory. We define two
different function calls for these two scenarios:
\begin{lstlisting}
int PtlHandlerPutFromHost(ptl_size_t offset, ptl_size_t len, ptl_req_ct_t ct, ptl_ack_req_t ack_req, ptl_process_t target_id, ptl_match_bits_t match_bits, ptl_size_t remote_offset, void* user_ptr, ptl_hdr_data_t hdr_data);
\end{lstlisting}
This call enqueues a put operation from host memory. This operation shall
behave as if it was posted by the host. The offset is relative to the
ME, other fields such as pt\_index, eq\_hdl, etc. are inherited from ME.
The call simply enqueues an operation and may thus not block.
\begin{lstlisting}
int PtlHandlerPutFromDevice(const void *local, ptl_size_t len, ptl_req_ct_t ct, ptl_ack_req_t ack_req, ptl_process_t target_id, ptl_match_bits_t match_bits, ptl_size_t remote_offset, void* user_ptr, ptl_hdr_data_t hdr_data);
\end{lstlisting}
This call performs a (single-packet) put operation from device memory.
The data is sent from HPU memory and len must be at most
max\_payload\_size, other fields such as pt\_index, eq\_hdl, etc. are
inherited from ME. This function may block until it is completed.
Similar handler functions are specified for \lstinline$PtlHandlerGet*$ and
\lstinline$PtlHandlerAtomic*$. 

\hpara{counter manipulation}
Counters can be manipulated with the following calls:
\begin{lstlisting}
int PtlHandlerCTInc(ptl_ct_event_t increment);
int PtlHandlerCTGet(ptl_ct_event_t *event);
int PtlHandlerCTSet(ptl_ct_event_t new_ct);
\end{lstlisting}
All three of them atomically read or modify a counter with the expected
semantics.

\section{Computational Results Analysis: sPIN: High-performance
streaming Processing in the Network}

\subsection{Abstract}

To increase the trust in our simulation results, we now show details of
how we implemented the simulation. Our system allows us to run real handler
C codes with only trivial changes in the LogGOPSim/gem5 environment.
The handlers run in a gem5 simulation and are compiled with a real
gcc cross-compilation tool chain. Because the handlers are relatively
simple, we present the complete source code here so the reader can
validate the functionality and get a feeling of the low complexity and
high power of the overall approach. 

We also provide some selected trace diagrams that our simulator can
produce. These diagrams help understanding how time is spent in the
different systems and can lead to additional insights beyond the
explanations in the paper. They are neither post-processed nor intended
for the main paper, we apologize for the rendering complexity due to the
simple output interface. Yet, we believe that, if they are carefully
analyzed, they increase the reader's confidence in the simulation
correctness as they provide more intuition of the main benefits of
\abbr{} (packetized pipelining and NIC-side processing).

Both, the source code and the diagrams can be found in the package to
reproduce the experiments. We show them here simply for the convenience
of the reader.

\subsection{Results Analysis Discussion}

As any simulation approach, we made several assumptions about the speeds
of the future devices. They are all covered in the available source code
and described in the paper. We believe that our simulations accurately
reflect reality and a real system could be built to achieve similar
results. Both simulators have been calibrated for the target system.
LogGOPSim is calibrated for a modern InfiniBand system and gem5 is
calibrated to a modern ARM architecture (see references to the reports
in the main document).

\subsection{Summary: Handler code used in this paper}\label{app:code}

We now present the raw C source codes and some selected visualizations
to understand the system better.

\subsubsection{Ping pong}\label{sec:pingpongcode}
~\\

\begin{lstlisting}
#define PTL_MAX_SIZE 4096
#define STREAMING 1

typedef struct {
  ptl_size_t offset;
  ptl_process_t source;
  ptl_size_t length;
  bool stream;
} pingpong_info_t;

int pingpong_header_handler(const ptl_header_t h, void *state) {
  pingpong_info_t *i = state;

  if(h.length > PTL_MAX_SIZE || !STREAMING) {
    i->stream = false;
    i->length = h.length;
    return PROCEED; // don't execute any other handlers
  } else {
    i->source = h.source_id;
    i->stream = true;
    return PROCESS_DATA; // execute payload handler to put from device
  }
}

int pingpong_payload_handler(const ptl_payload_t p, void *state) {
  pingpong_info_t *i = state;

  PtlHandlerPutFromDevice(p.base, p.length, 1, 0, i->source, 10, 0, NULL, 0);

  return SUCCESS;
}

int pingpong_completion_handler(int dropped_bytes, bool flow_control_triggered, void *state) {
  pingpong_info_t *i = state;
  
  if(!i->stream) PtlHandlerPutFromHost(i->offset, i->length, 1, 0, i->source, 10, 0, NULL, 0);

  return SUCCESS;
}


\end{lstlisting}

The following figure shows the trace simulation output for RDMA 
8 KiB ping-pong. The red and blue sections at rank 0 and 1 indicate message
posting overheads and the vertical blue bars represent data transfers. 
\includegraphics[width=\columnwidth]{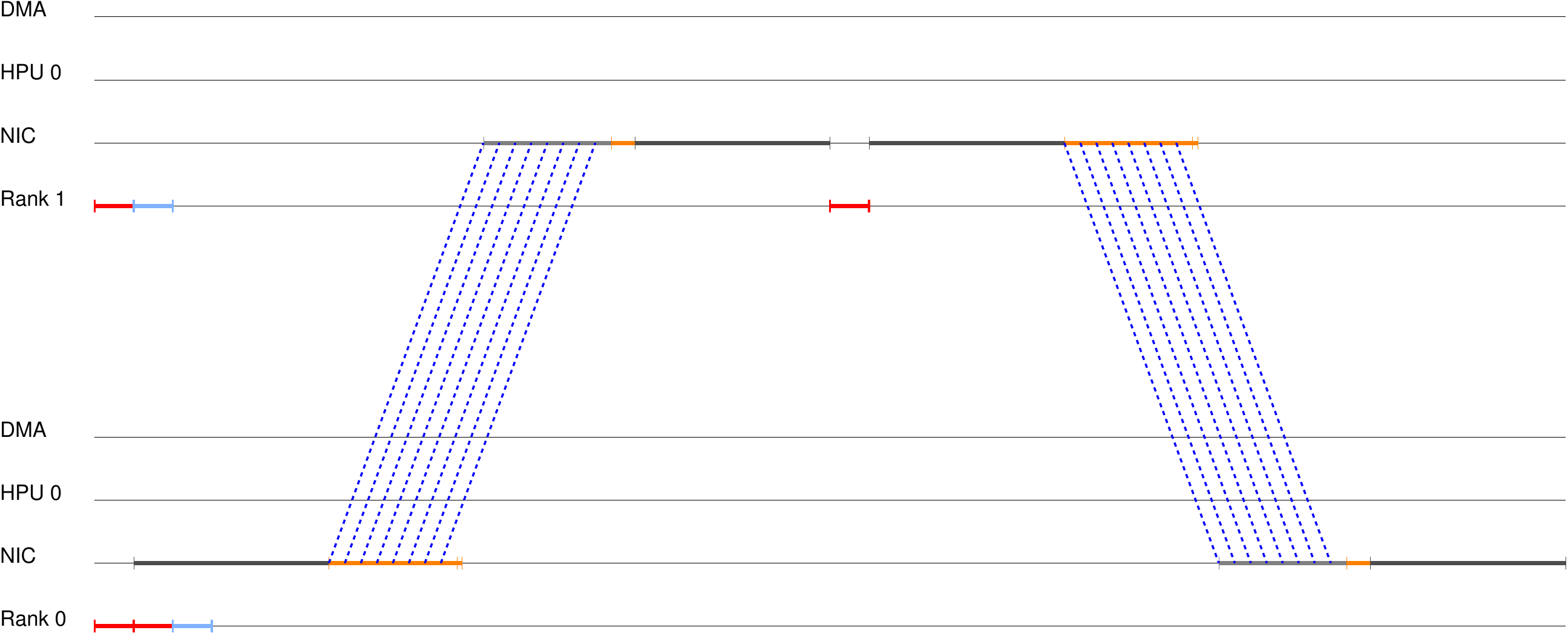}

The next figure shows the trace simulation output for \abbr{} store
mode. Here, the two packets are matched separately and the completion
handler is executed on HPU 0.
\includegraphics[width=\columnwidth]{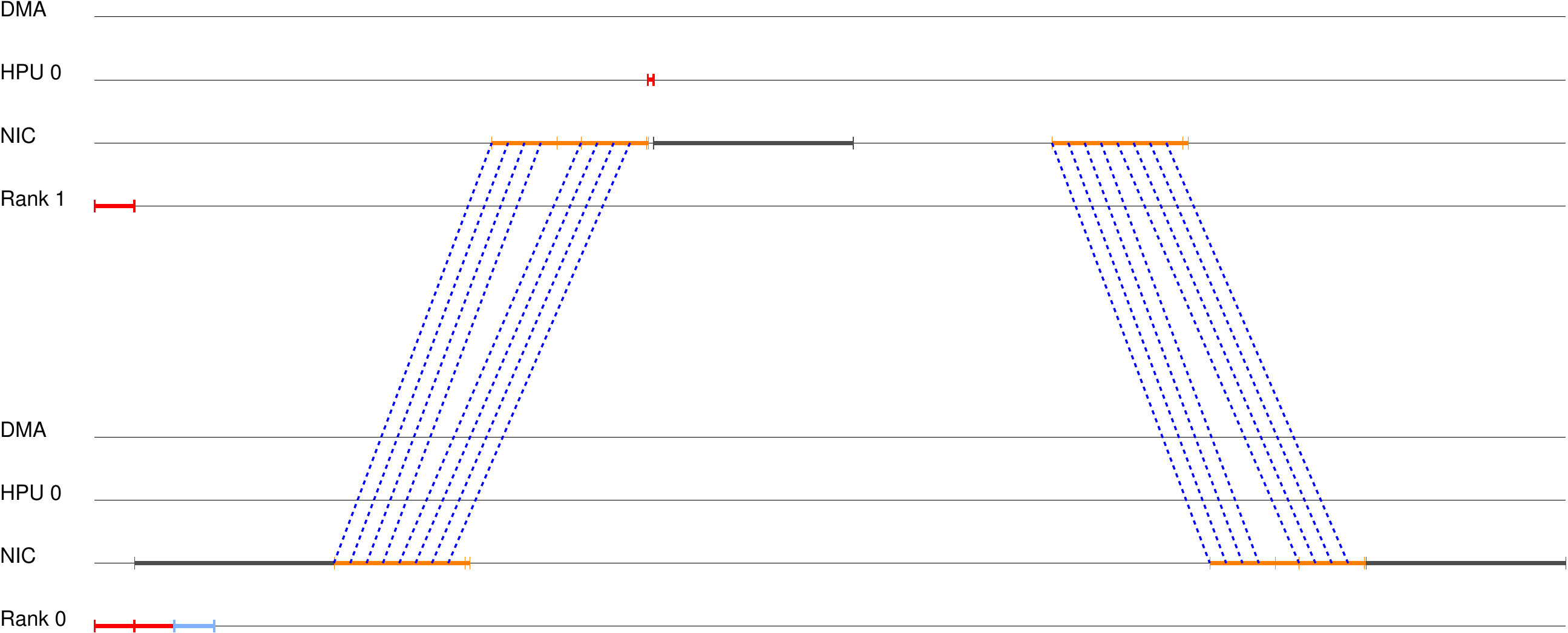}

The next figure shows the trace simulation output for \abbr{} stream
mode. Here, the two packets are matched separately and two instances of
the payload handler are executed on HPU 2 and HPU 0, respectively. The
first reply packet is already sent before the second is fully received (the
transceiver is full duplex).
\includegraphics[width=\columnwidth]{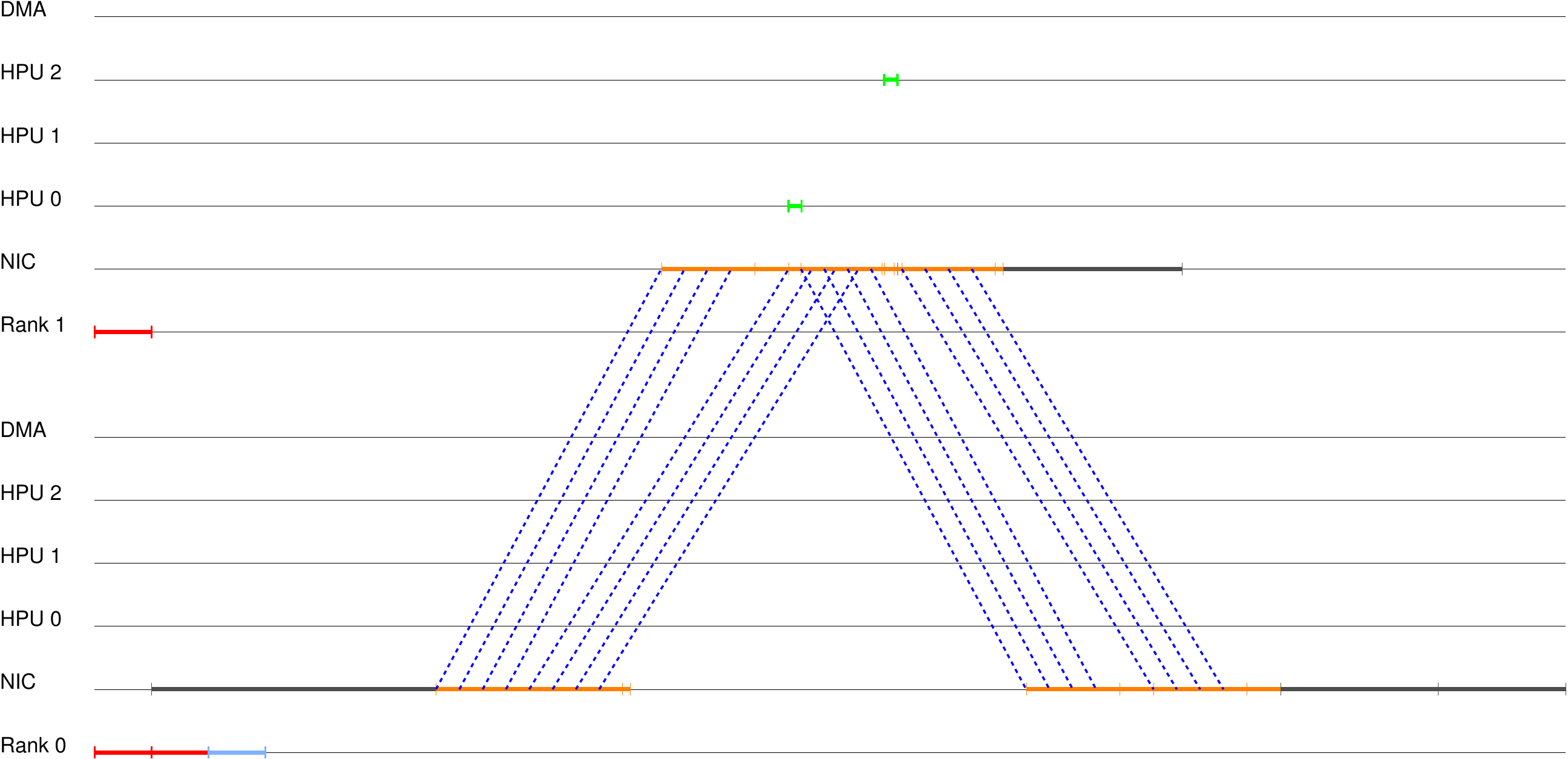}

\subsubsection{Accumulate}\label{sec:accumulatecode}
~\\

\begin{lstlisting}
typedef struct {
  ptl_size_t offset;
  bool pong;
  ptl_process_t source;
} accumulate_info_t;


int accumulate_header_handler(const ptl_header_t h, void *state) {
  accumulate_info_t *i = state;
  if(i->pong) i->source = h.source_id;
  return PROCESS_DATA; 
}


int accumulate_payload_handler(const ptl_payload_t p, void * state) {
  accumulate_info_t *i = (accumulate_info_t*)state; 

  float* buf = malloc(p.length/sizeof(float));
  PtlHandlerDMAFromHostB(i->offset+p.offset, buf,  p.length, PTL_ME_HOST_MEM); 
 
  float *data = (float*)p.base;

  for(int j=0; j<p.length/sizeof(float); j+=2) {	
    buf[j] = data[j]*buf[j] - data[j+1]*buf[j+1];
    buf[j+1] = data[j]*buf[j+1] - data[j+1]*buf[j];	
  }

  PtlHandlerDMAToHostB(buf, i->offset+p.offset, p.length, PTL_ME_HOST_MEM);
  if(i->pong) PtlHandlerPutFromDevice(buf, p.length, 1, 0, i->source, 10, 0, NULL, 0);

  free(buf);
  return SUCCESS;
}

\end{lstlisting}

The following figure shows the trace simulation output for RDMA 
accumulate of 8KiB. The elements are similar to ping pong but the
accumulate is performed at rank 1's CPU (multi-core not displayed here). 
\includegraphics[width=\columnwidth]{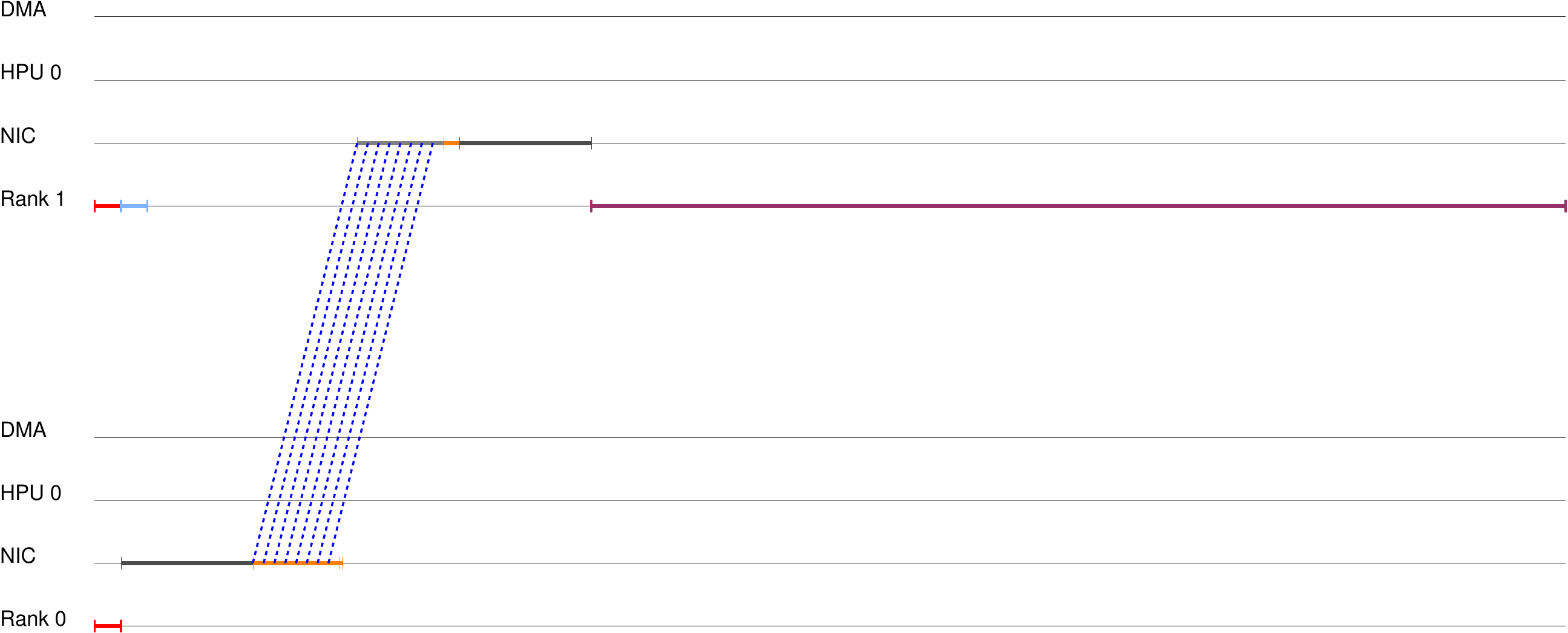}

The next figure shows the trace simulation output for \abbr{} store
mode. The simulation schedules the two packets for HPU 0 and HPU 2,
respectively. The HPUs issue competing DMA requests to the host memory.
The DMA sections on the figure depict the blocking time of the handler.
In other words, the time spent by the HPU to complete the DMA request. Thus,
DMA requests can be overlapped on the generated diagrams.  Long sections
stand for DMAFromHost, since we pay two DMA latencies to read the data,
and short ones for DMAToHost, which blocks HPUs to initiate the write
request. 
\includegraphics[width=\columnwidth]{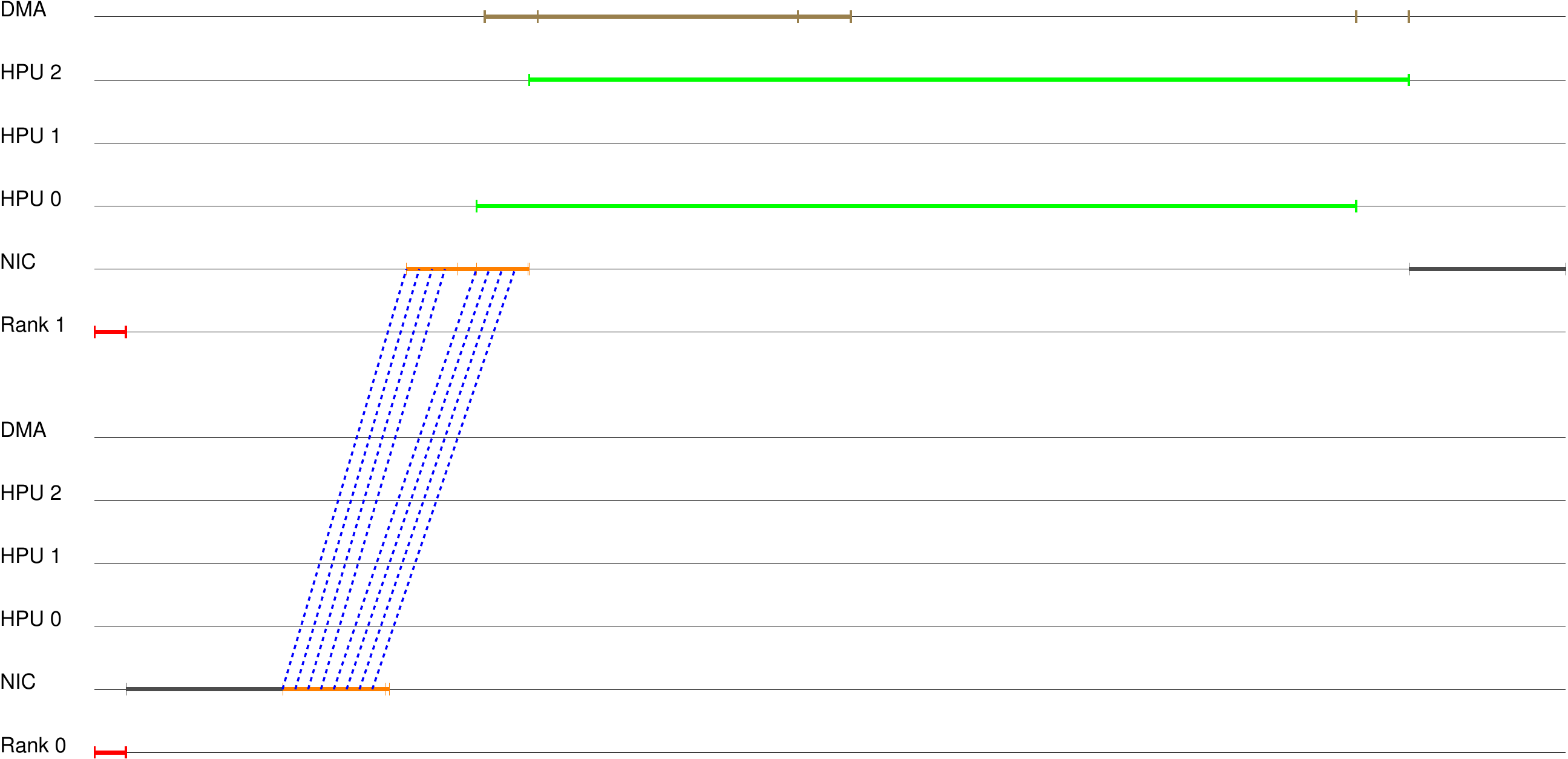}

\subsubsection{Broadcast}\label{sec:broadcastcode}
~\\

\begin{lstlisting}
#define PTL_MAX_SIZE 4096
#define STREAMING 1

typedef struct {
  ptl_size_t offset;
  ptl_process_t my_rank;
  ptl_process_t p;           // total number of hosts
  bool stream;
  ptl_size_t length;
} bcast_info_t;


int bcast_header_handler(const ptl_header_t h, void *state) {
  bcast_info_t *i = state;

  if(h.length > PTL_MAX_SIZE || !STREAMING) {
    i->stream = false;
    i->length = h.length;
    return PROCEED; // don't execute any other handlers
  } else {
    i->stream = true;
    return PROCESS_DATA; // execute payload handler to put from device
  }
}

int bcast_payload_handler(const ptl_payload_t p, void *state) {
  bcast_info_t *i = state;

  for (uint32_t half = i->p/2; half>=1; half/=2 )
    if (i->my_rank % (half*2) == 0)
      PtlHandlerPutFromDevice(p.base, p.length, PTL_LOCAL_ME_CT, 0, i->my_rank+half, 10, 0, NULL, 0);
	
  return SUCCESS;
}

int bcast_completion_handler(int dropped_bytes, bool flow_control_triggered, void *state) {
  bcast_info_t *i = state;
  
  if(!i->stream)
    for (uint32_t half = i->p/2; half>=1; half/=2 )
      if (i->my_rank % (half*2) == 0)
        PtlHandlerPutFromHost(i->offset, i->length, PTL_LOCAL_ME_CT, 0, i->my_rank+half, 10, 0, NULL, 0);
	
  return SUCCESS;
}


\end{lstlisting}

The following figure shows the trace simulation output for RDMA 
broadcast of 8KiB to 16 ranks. The elements are similar to ping pong and
rank 0 is the root. 
\includegraphics[width=\columnwidth]{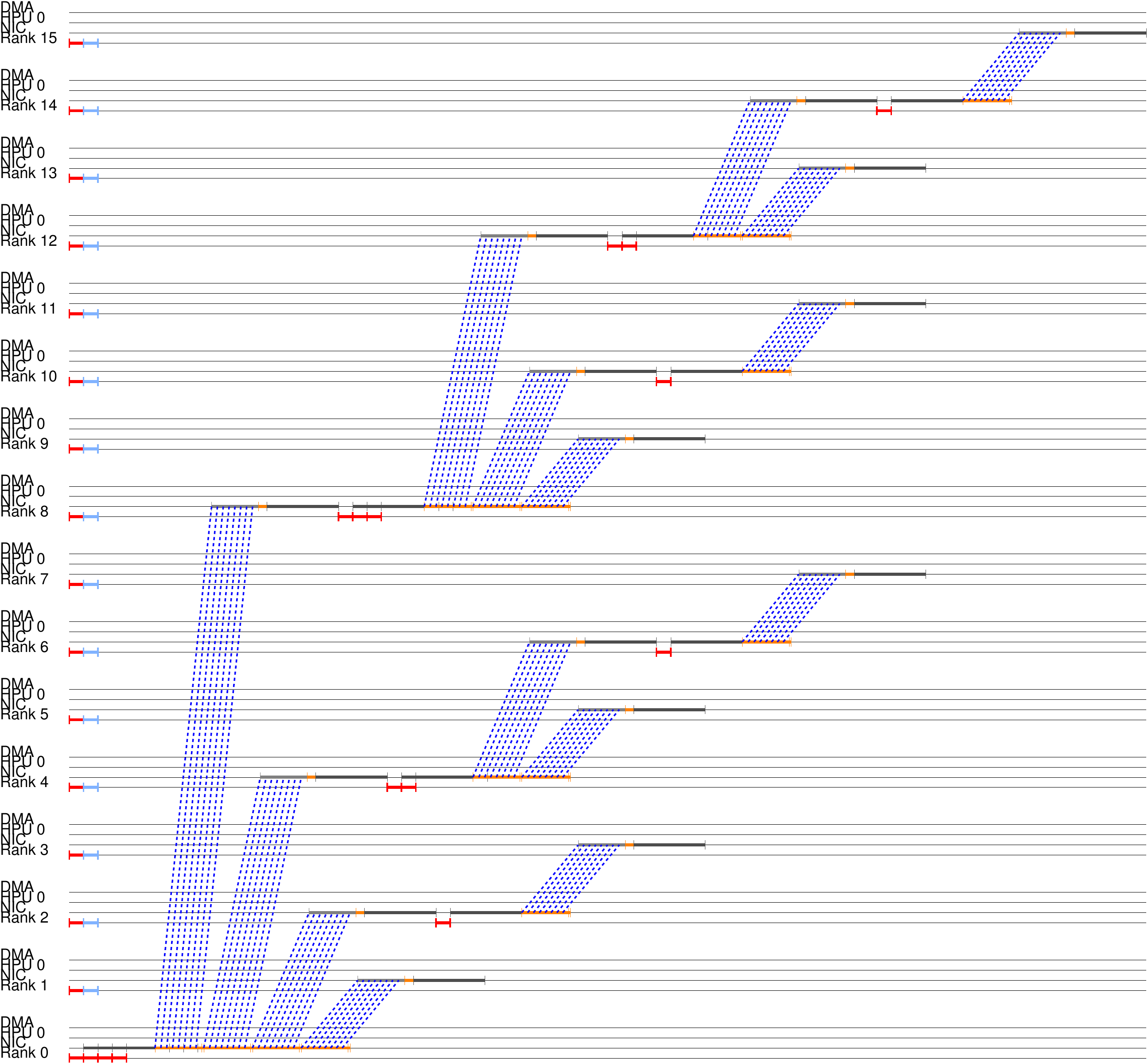}

The next figure shows the trace simulation output for the \abbr{}
broadcast. We see again how the packets are forwarded in a pipelined
manner by the HPUs at each node. The first packets are sent before the
message is fully received, illustrating the wormhole-routing-like
behavior. 
\includegraphics[width=\columnwidth]{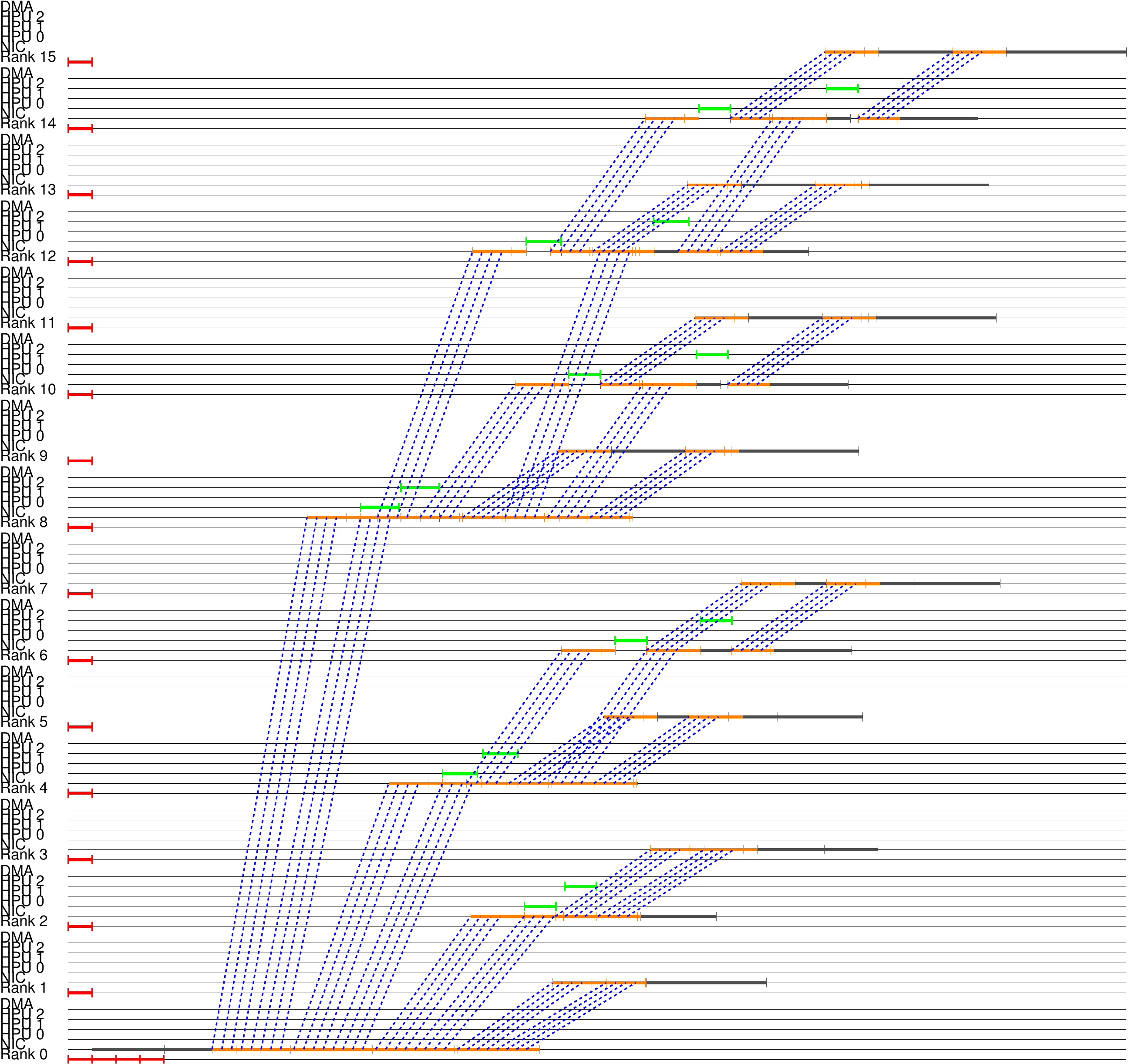}

\subsubsection{Strided Datatype}\label{sec:datatypecode}
~\\

\begin{lstlisting}
typedef struct {
  ptl_size_t offset;
  ptl_size_t vlen;
  ptl_size_t stride;
} ddtvec_info_t;


int ddtvec_payload_handler(const ptl_payload_t p, void *state) {
  ddtvec_info_t *i = (ddtvec_info_t*)state; 
  ptl_size_t first_seg_num = (i->offset+p.offset)/i->vlen; // rounded down implicitly
  ptl_size_t last_seg_num = (i->offset+p.offset+p.length)/i->vlen; // rounded down implicitly

  int offset_in_packet = 0;
  for(int j = first_seg_num; j<=last_seg_num; j++)
  {
        ptl_size_t offset_in_block = (i->offset+p.offset+offset_in_packet)%i->vlen;  
	ptl_size_t offset =  j * (i->vlen+i->stride) + offset_in_block;
        ptl_size_t size =  (i->vlen - offset_in_block) < (p.length - offset_in_packet) ? (i->vlen - offset_in_block) : (p.length - offset_in_packet) ;
	PtlHandlerDMAToHostB(p.base+offset_in_packet, offset, size, PTL_ME_HOST_MEM); 
	offset_in_packet+=size;
  }
 
  return SUCCESS;
}
\end{lstlisting}

The following figure shows the trace simulation output for RDMA
reception of a datatype with 32 8 KiB blocks (256 KiB total). The
elements are similar to ping pong and the unpack is performed by the
CPU. 
\includegraphics[width=\columnwidth]{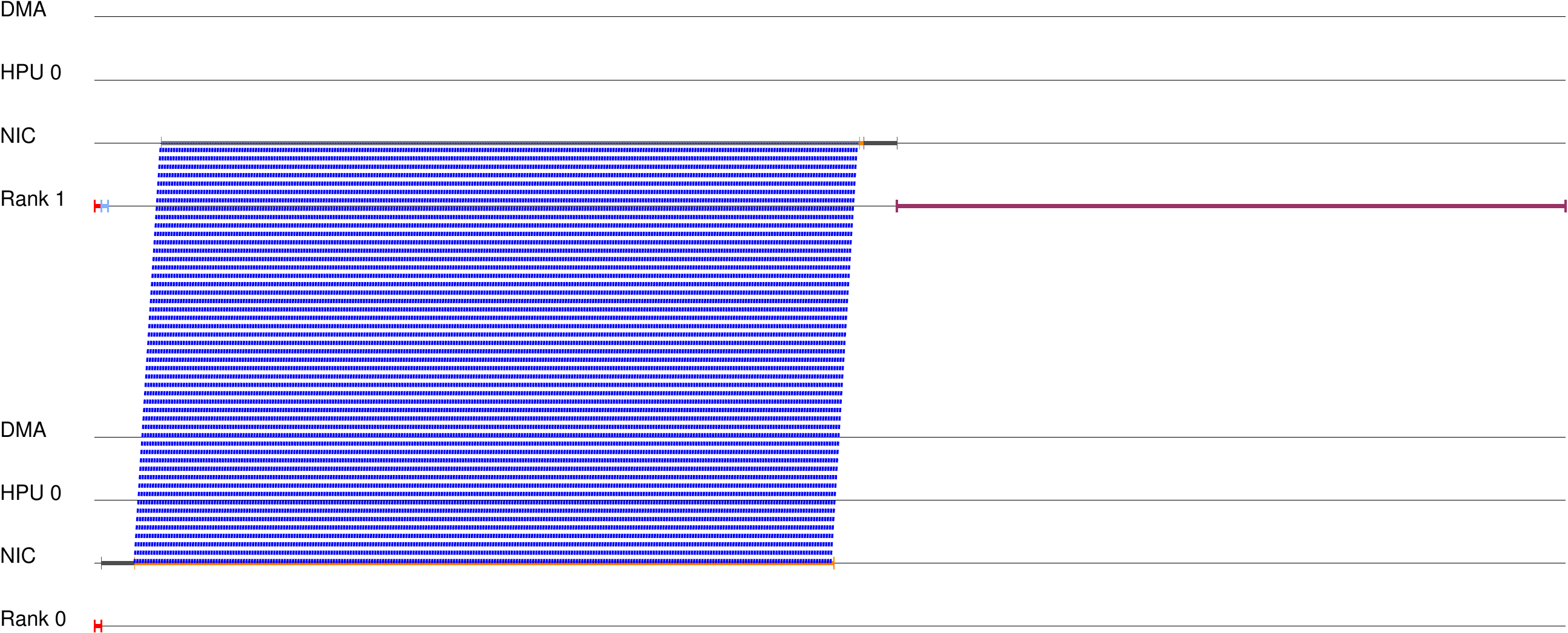}

The next figure shows the trace simulation output for the \abbr{}
datatype receive. Packets are blocked and each block is processed by one
of the four HPUs, which issues a local DMA transaction to host memory. For each packet the payload handler was executed (64 calls overall), which deposited data directly to right locations in memory, whereas  RDMA system should wait for the arrival of the whole message to process it. 
\includegraphics[width=\columnwidth]{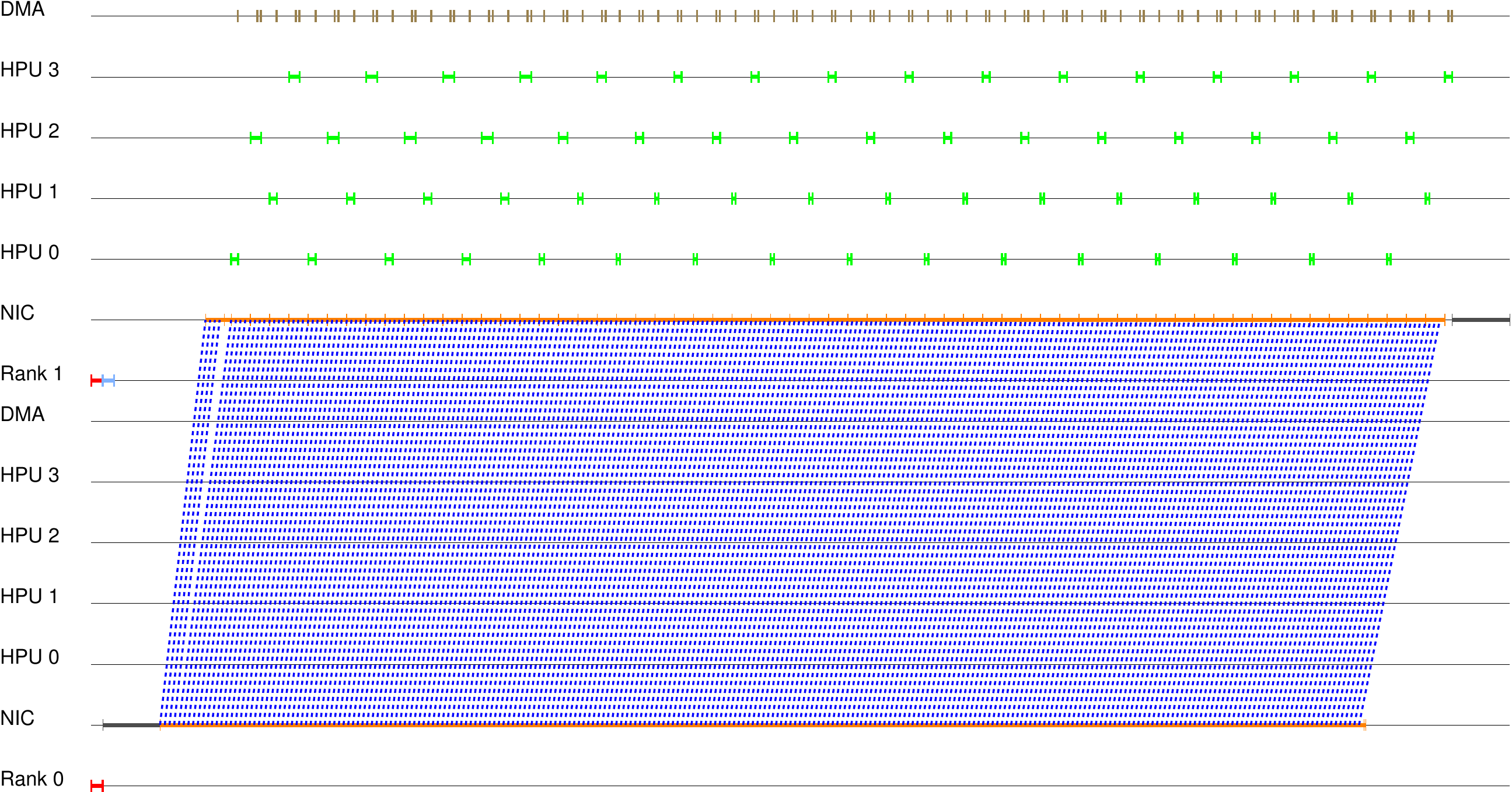}

\subsubsection{Reed-Solomon}\label{sec:reedsolomoncode}
~\\

\begin{lstlisting}
#define PARITY_TAG 53

/* Code for data server */

typedef struct {
  ptl_process_t source;
  ptl_process_t parity;
  ptl_size_t offset;
} primary_info_t;

typedef struct {
  int length;
} ptl_user_header_t;

int primary_write_header_handler(const ptl_header_t h, void * state) {
  primary_info_t *i = state;
  i->source=h.source_id;
  return PROCESS_DATA;
}

int primary_write_payload_handler(const ptl_payload_t p, void * state) {
  primary_info_t *i = state; 

  uint32_t* buf = malloc(p.length/sizeof(uint32_t));
  PtlHandlerDMAFromHostB(i->offset+p.offset, buf,  p.length, PTL_ME_HOST_MEM);
 
  uint32_t *data = (uint32_t*)p.base;
 
  for(int i=0; i<p.length/sizeof(uint32_t); i++) 
	buf[i] = (buf[i] ^ data[i]); 


  PtlHandlerDMAToHostB(buf, i->offset+p.offset, p.length, PTL_ME_HOST_MEM); 
  PtlHandlerPutFromDevice(buf, p.length, PTL_LOCAL_ME_CT, 0, i->parity, PARITY_TAG, 0, NULL, i->source);

  free(buf);
  return SUCCESS;
}

int primary_read_header_handler(const ptl_header_t h, void * state) {
  primary_info_t *i = state; 
  PtlHandlerPutFromHost(h.offset, h.user_hdr.length, PTL_LOCAL_ME_CT, 0, h.source_id, h.match_bits, 0, NULL, 0);
  return PROCEED;
}

int primary_send_acknowledgement_header_handler(const ptl_header_t h, void * state) {
  uint8_t reply = SUCCESS;
  PtlHandlerPutFromDevice(&reply, 1, PTL_LOCAL_ME_CT, 0, h.user_hdr.client, h.match_bits, 0, NULL, 0);
  return PROCEED;
}


/* Code for parity server */

typedef struct {
  ptl_process_t source;
  ptl_process_t client;
  ptl_size_t offset;
} parity_info_t;

typedef struct {
  int client;
} ptl_user_header_t;

int parity_update_header_handler(const ptl_header_t h, void * state) {
  parity_info_t *i = state;
  i->source=h.source_id;
  i->client=h.user_hdr.client;
  return PROCESS_DATA;
}

int parity_update_payload_handler(const ptl_payload_t p, void * state)  {
  parity_info_t *i = state;
  uint32_t* buf = malloc(p.length/sizeof(uint32_t));

  PtlHandlerDMAFromHostB(i->offset+p.offset, buf,  p.length, PTL_ME_HOST_MEM);

  uint32_t *data = (uint32_t*)p.base;
 
  for(int i=0; i<p.length/sizeof(uint32_t); i++)
	buf[i] = (buf[i] ^ data[i]); 

  PtlHandlerDMAToHostB(buf, i->offset+p.offset, p.length, PTL_ME_HOST_MEM); 
  free(buf);
  return SUCCESS;
}

int parity_update_completion_handler(int dropped_bytes, bool flow_control_triggered, void *state)  {
  parity_info_t *i = state;
  uint8_t reply = SUCCESS;
  PtlHandlerPutFromDevice(&reply,1, PTL_LOCAL_ME_CT, 0, i->source, 30, 0, NULL, i->client);
  return SUCCESS;
}

\end{lstlisting}

The following figure shows the trace simulation output for RDMA write
using Reed-Solomon coding in a RAID-5 configuration. Here, rank 0 is the
client and it sends updates to all four servers (rank 2-5), which then
update the parity at rank 1 and acknowledge back to rank 0. P4 case is different from RDMA, since the data server can predict getting acknowledgment from parity server and initiate triggered put right after the sending the data to it in order to reply the client faster. 
\includegraphics[width=\columnwidth]{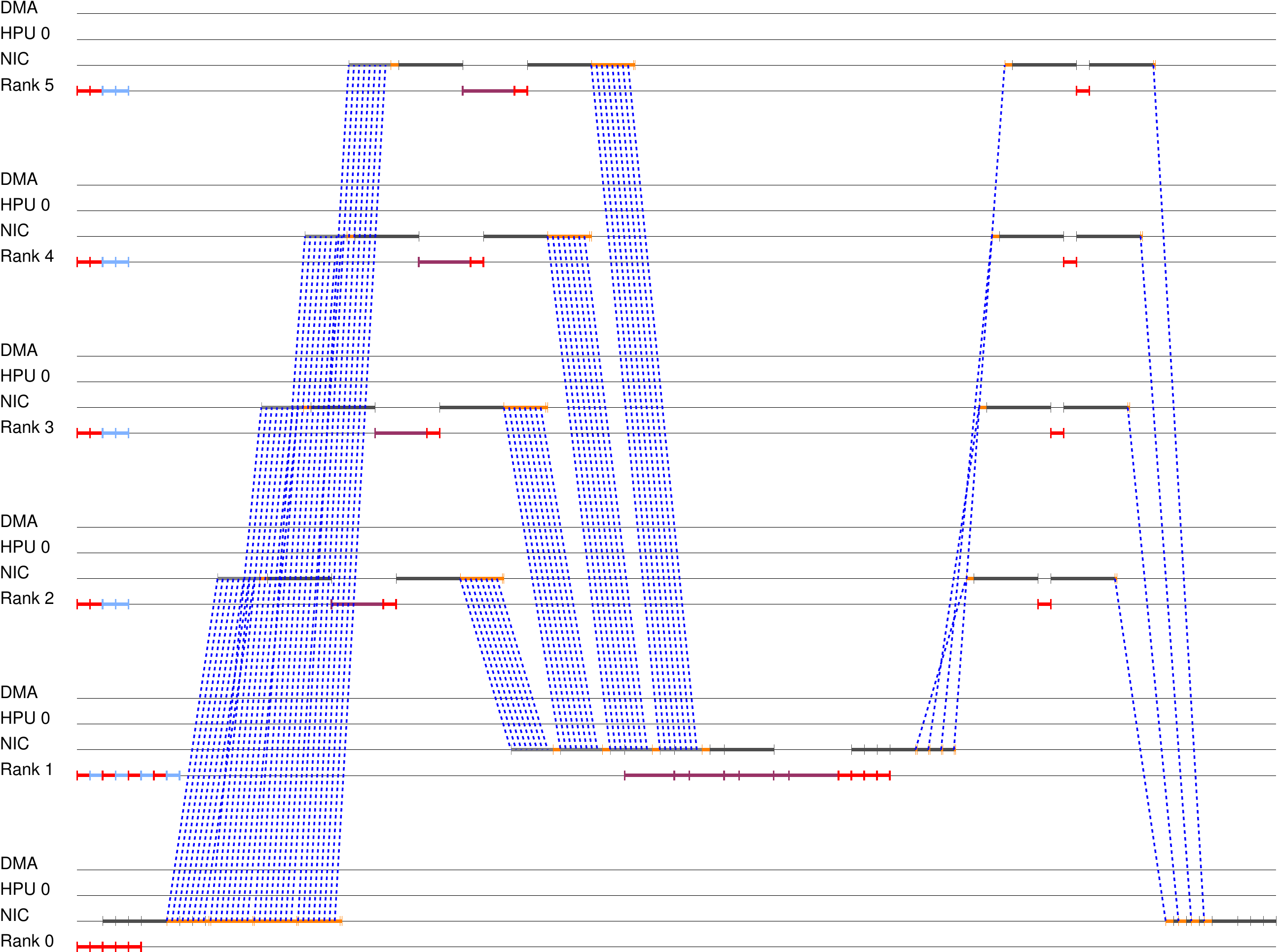}

The next figure shows the trace simulation output for the \abbr{}
RAID-5 update. Again, packets applied to the local host memory via DMA
and they are pipelined simultaneously through the network towards the
parity rank 1. Also acknowledgments are sent directly from the NICs.
\includegraphics[width=\columnwidth]{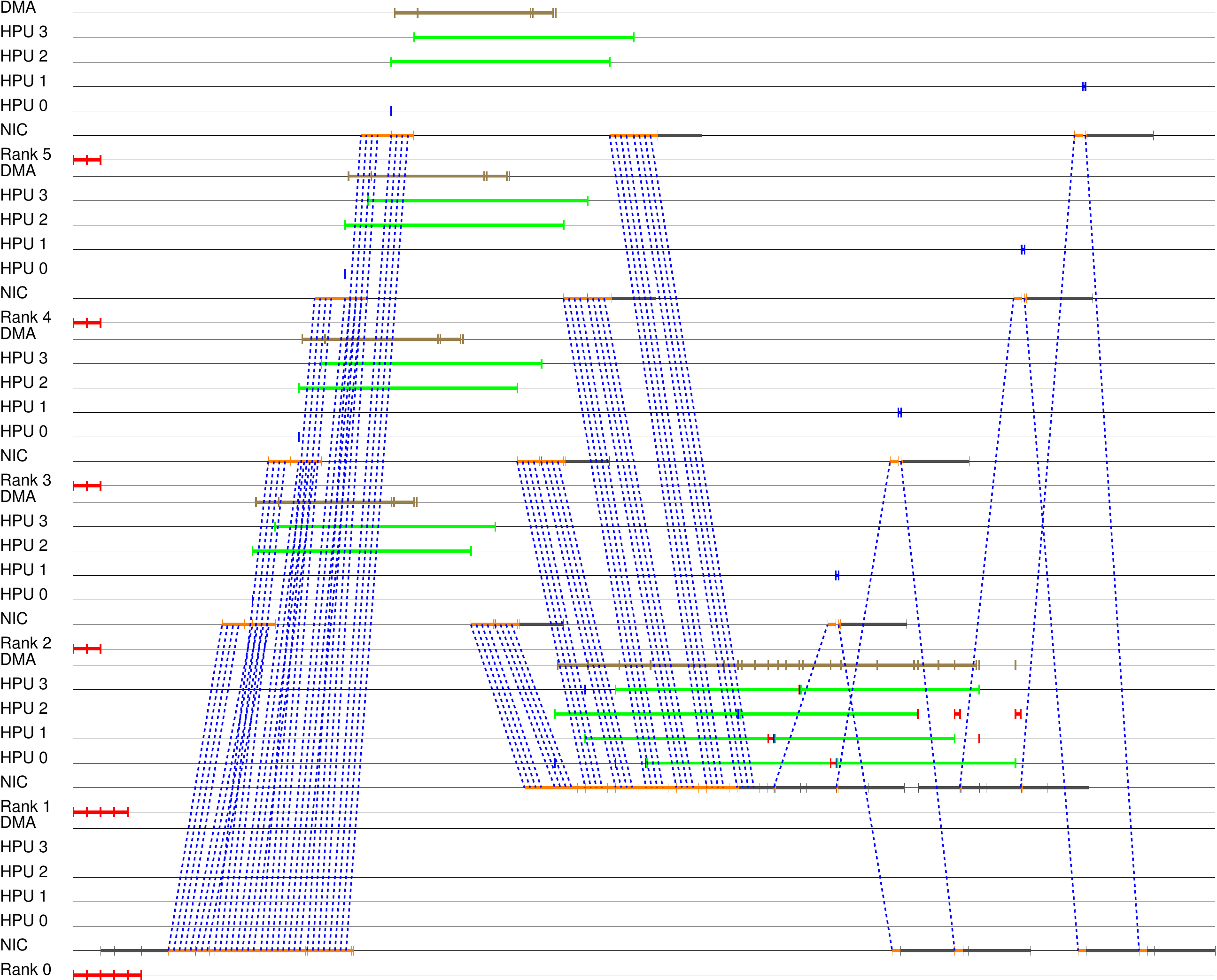}

\end{document}